\documentclass[usenatbib]{mnras}
\usepackage{amsmath}
\usepackage{amssymb}
\usepackage{graphicx}
\usepackage[utf8]{inputenc}
\usepackage{listings}
\usepackage{geometry}
\usepackage[normalem]{ulem}
\usepackage{soul}
\usepackage[dvipsnames]{xcolor}
\usepackage{subcaption}
\usepackage{cleveref}

\newcommand{\equ}[1]{eq.~(\ref{eq:#1})}
\newcommand{\equs}[1]{eqs.~(\ref{eq:#1})}
\newcommand{\equm}[1]{(\ref{eq:#1})}

\newcommand{\equnp}[1]{eq.~\ref{eq:#1}}
\newcommand{\equsnp}[1]{eqs.~\ref{eq:#1}}
\newcommand{\equmnp}[1]{\ref{eq:#1}}

\newcommand{\fig}[1]{Fig.~\ref{fig:#1}}

\newcommand{\be}{\begin{equation}}
\newcommand{\ee}{\end{equation}}
\newcommand{\bea}{\begin{eqnarray}}
\newcommand{\eea}{\end{eqnarray}}

\newcommand{\no}{\noindent}

\newcommand{\msun}{{\rm M}_\odot}
\newcommand{\Msun}{M_\odot}

\newcommand{\ifm}[1]{\relax\ifmmode#1\else$\mathsurround=0pt #1$\fi}
\newcommand{\kms}{\ifmmode\,{\rm km}\,{\rm s}^{-1}\else km$\,$s$^{-1}$\fi}

\newcommand{\kpc}{\,{\rm kpc}}
\newcommand{\pc}{\,{\rm pc}}
\newcommand{\Gyr}{\,{\rm Gyr}}
\newcommand{\yr}{\,{\rm yr}}

\newcommand{\K}{\,{\rm K}}

\newcommand{\ltsima}{$\; \buildrel < \over \sim \;$}
\newcommand{\lsim}{\lower.5ex\hbox{\ltsima}}
\newcommand{\gtsima}{$\; \buildrel > \over \sim \;$}
\newcommand{\gsim}{\lower.5ex\hbox{\gtsima}}

\newcommand{\rmd}{\rm d}

\newcommand{\rstream}{R}
\newcommand{\rhalo}{r}
\def\Rso{R_{\rm s,0}}

\def\sy{\,M_\odot\, {\rm yr}^{-1}}

\def\cmc{\,{\rm cm}^{-3}}

\def\M*{M_{\rm *}}

\def\Mv{M_{\rm v}}
\def\Rv{R_{\rm v}}
\def\Vv{V_{\rm v}}

\def\Rs{R_{\rm s}}

\def\rhob{\rho_{\rm b}}
\def\rhos{\rho_{\rm s}}

\def\cb{c_{\rm b}}
\def\cs{c_{\rm s}}

\def\Vs{V_{\rm s}}

\def\tsc{t_{\rm sc}}
\def\tent{t_{\rm ent}}
\def\Mb{M_{\rm b}}

\def\Pi{\varpi_{_{\rm I}}}
\def\rmd{{\rm d}}

\defcitealias{M16}{M16}
\defcitealias{P18}{P18}
\defcitealias{M18b}{M19}
\defcitealias{M20a}{M20a}
\defcitealias{M20b}{M20b}

\usepackage{color}

\begin{document} 

\large
\title[Entrainment of the Hot CGM onto Cold Streams]
{Entrainment of Hot Gas into Cold Streams: The Origin of Excessive Star-formation Rates at Cosmic Noon}

\author[Aung et al.] 
{\parbox[t]{\textwidth} 
{
Han Aung$^1$\thanks{E-mail: han.aung@mail.huji.ac.il},
Nir Mandelker$^{1}$,
Avishai Dekel$^{1,2}$,
Daisuke Nagai$^{3}$,
Vadim Semenov$^{4}$,
Frank C. van den Bosch$^{5}$
} 
\\ \\ 
$^1$Centre for Astrophysics and Planetary Science, Racah Institute of Physics, The Hebrew University, Jerusalem 91904, Israel \\
$^2$SCIPP, University of California, Santa Cruz, CA 95064, USA\\
$^3$Department of Physics, Yale University, New Haven, CT 06520, USA \\
$^4$Center for Astrophysics, Harvard \& Smithsonian, 60 Garden St, Cambridge, MA 02138, USA\\
$^5$Department of Astronomy, Yale University, PO Box 208101, New Haven, CT 06520, USA
}
\date{} 

\maketitle 

\label{firstpage} 

\begin{abstract}
We explore the evolution of cold streams from the cosmic web that feed galaxies through their shock-heated circumgalactic medium (CGM) at cosmic noon, $z\simeq 1-5$. In addition to the hydrodynamical instabilities and radiative cooling that we have incorporated in earlier works, we embed the stream and the hot CGM in the gravitational potential of the host dark-matter halo, deriving equilibrium profiles for both. Self-gravity within the stream is tentatively ignored. 
We find that the cold streams gradually entrain a large mass of initially hot CGM gas that cools in the mixing layer and condenses onto the stream. This entrainment, combined with the acceleration down the gravitational potential well, typically triples the inward cold inflow rate into the central galaxy, compared to the original rate at the virial radius, which makes the entrained gas the dominant source of gas supply to the galaxy.
The potential sources for the hot gas to be entrained are recycled enriched gas that has been previously ejected from the galaxy, and fresh virial-shock-heated gas that has accumulated in the CGM.
This can naturally elevate the star formation rate in the galaxy by a factor of $\sim 3$ compared to the gas accretion rate onto the halo, thus explaining the otherwise puzzling observed excess of star formation at cosmic noon. When accounting for self-shielding of dense gas from the UV background, we find that the energy radiated from the streams, originating predominantly from the cooling of the entrained gas, is consistent with observed Lyman-$\alpha$ blobs around galaxies.
\end{abstract}



\section{Introduction}
\label{sec:intro}
The large-scale structure of the Universe is dominated by filamentary structures that span several Mpc \citep{Zeldovich70,Bond96,Springel05}. Intergalactic gas cools and condenses toward the centres of these dark matter filaments, forming a network of baryon-dominated intergalactic gas streams \citep{db06,Birnboim16, Lu2023}. These become sites of active galaxy formation, evident in the observed distributions of galaxies that closely mimic the predicted ``cosmic web''. \citep[e.g.,][]{Tegmark04,Huchra05,Burchett.etal.20}. 

\smallskip
At the nodes of the cosmic web, the most massive haloes reside at the intersection of several filaments and are penetrated by gas streams flowing along them. These streams constitute the primary mode of gas accretion onto massive galaxies with baryonic mass $\gtrsim 10^{11}\Msun$ \citep{Keres05,Dekel09,Danovich12,Zinger16}. At redshifts $z\gtrsim 2$, simulations suggest that streams feeding galactic haloes remain dense and cold, with temperatures of $\sim 10^4\K$, as they travel through the hot circumgalactic medium (CGM, the gas outside galaxies but within dark matter haloes) towards the central galaxy (\citealp{Keres05,db06,Ocvirk08,Dekel09,CDB,FG11,vdv11}, though see also \citealp{Nelson13,Nelson16}). The filamentary structure in such systems can thus be maintained down to scales of tens of $\kpc$ around galaxies (though see below), where it has been suggested that they may fragment due to gravitational instability \citep[hereafter GI;][]{DSC,Genel12,M18a,A19}. Although these cold circumgalactic streams are difficult to detect directly, recent observations have revealed massive extended cold components in the CGM of high redshift galaxies, whose spatial and kinematic properties are consistent with predictions for cold streams \citep{Bouche13,Bouche16,Prochaska14,Cantalupo14,Martin14a,Martin14b,Martin19,Borisova16,Fumagalli17,Leclercq17,Arrigoni18, Daddi.etal.21, Emonts2023, Zhang2023}. 

\smallskip
As the cold streams flow through the hot CGM, they are subject to Kelvin-Helmholtz Instability (KHI) due to the strong shear between the inflowing streams and the hot CGM. Correctly modelling KHI in cosmological simulations of galaxies requires exquisite resolution in the streams, particularly in the mixing layer that forms between the stream and the CGM. However, current state-of-the-art simulations typically adopt a (quasi-)Lagrangian refinement strategy, where the mass of each resolution element is roughly constant. 
This leads to poor spatial resolution in low-density regions, such as streams in the outer CGM. Thus, different simulation methods lead to different conclusions for cold streams, some simulations predicting that streams are disrupted at $\gsim 0.5\Rv$, with $\Rv$ the halo virial radius \citep{Nelson13}, and others predicting that streams remain cold and dense down to $\lsim 0.25\Rv$ \citep{Keres05,FG10,CDB,Danovich15}.

\smallskip
KHI in cold accretion streams has been extensively studied using idealised simulations of cylindrical streams, with over 100 resolution elements across the stream, sufficient to resolve the linear and non-linear evolution of KHI in the stream. 
In a series of papers, \citet{M16,M18b} and \citet{P18} (hereafter \citetalias{M16, M18b,P18}, respectively) presented a detailed study of KHI in a dense supersonic cylinder flowing through a hot, diffuse background, representing cold streams flowing through the shock-heated CGM surrounding massive high-z galaxies. They found that KHI at low Mach numbers manifests itself as turbulent eddies at the interface between the two fluids (surface modes), while at high Mach numbers KHI manifests itself as overstable waves reflected internally off the interface (body modes). In both cases, significant turbulent mixing occurs between the cold and hot fluids during the non-linear phase of the instability. This mixing drains momentum and energy from the cold component and can cause a total disruption of the stream in the CGM if $R_{\rm s}/\Rv\lesssim 0.05$, with $R_{\rm s}$ the stream radius and $\Rv$ the halo virial radius (M19). We elaborate on these studies in \cref{sec:khi}. 

\smallskip
\citet{A19} showed that self-gravity can counteract KHI but leads to gravitational instability at large values of the stream mass per unit length (hereafter line-mass), forming giant gas clumps. These clumps can potentially be sites for star-formation within streams outside the central galaxy \citep{M18a,Bennett.Sijacki.20}, and can also be important drivers of turbulence in the central galaxy \citep{DSC,Genel12,Ginzburg.etal.22} if they survive their transport to the centre of the host halo.

\smallskip
Radiative cooling also qualitatively affects the dynamic, morphological, and kinematic evolution of cold streams, which in turn profoundly impact the mass, structure, and kinematics of high-$z$ disc galaxies. \citet{M20a} (hereafter \citetalias{M20a}) showed that when the cooling in the turbulent mixing layer that forms at the interface between the stream and the CGM is more efficient than hydrodynamic mixing through KHI, the hot background gas cools and condenses onto the stream and becomes entrained in the flow. This causes the cold gas mass to increase, while its velocity and kinetic energy decrease.  
However, if cooling in the mixing layer is slower than hydrodynamic mixing, the stream will disrupt similarly to the no-cooling case. Similar conclusions have been reached for spherical clouds \citep[e.g.][]{GronkeOh2018,GronkeOh20,Li20,Sparre2020} and planar shear layers \citep[e.g.][]{Ji2019,Fielding20,Tan21}. 

\smallskip
The gravitational potential of the host halo can significantly alter the geometry and kinematics of the stream, as it causes streams to develop a conic shape, narrowing in size and increasing in density while accelerating towards the central galaxy. Several studies accounting for the halo potential in simulations \citep{Wang14, Hong2024} and in analytic models \citep[][hereafter \citetalias{M20b}]{M20b} show that the filaments feeding galaxies at cosmic noon are stable against a variety of instabilities. By modelling the properties of streams penetrating the virial shock as a function of halo mass and redshift, \citetalias{M20b} showed that most cosmological streams are expected to be in the fast cooling regime, where entrainment dominates. By further assuming that the streams and the CGM are in local pressure equilibrium at each halo-centric radius, and that the entrainment rates derived by \citetalias{M20a} could be applied at each radius using the local conditions in the streams and the CGM, we predicted that the entrainment only gets stronger deeper in the halo potential, and can lead to a net increase of up to a factor of 2 in the cold gas mass reaching the central galaxy compared to the cold gas mass entering the halo virial radius. Furthermore, the stream velocity increases due to the gravitational acceleration by the host-halo, albeit less than free-fall because of the deceleration induced by entrainment. Taken together, these results suggest that cold streams can supply the central galaxy directly with cold gas to fuel ongoing star-formation without being disrupted, potentially at even higher rates than the cold gas accretion rate onto the halo. In M20b, we also showed that entrainment results in enough Ly$\alpha$ emission outside $\sim 0.1\Rv$ to power observed Ly$\alpha$ blobs \citep{Steidel00,Steidel04,Steidel10,Matsuda06,Matsuda11,Cantalupo14,Martin14a,Martin14b,Martin19,Borisova16,Arrigoni18,Daddi.etal.21}, with predicted luminosities of $L_{\rm Ly\alpha}\sim (10^{42}-10^{44})\,{\rm erg\,s^{-1}}$ for halo masses of $\sim (10^{12}-10^{13})\msun$ at $z\sim (2-4)$.

\smallskip
While M20b predicted that streams continue to accelerate down the potential well until they reach the central galaxy, cold clouds flowing through a hot medium under an external gravitational potential have been shown to reach a terminal velocity due to strong braking caused by a hydrodynamic drag force along with entrainment \citep{Tan23}. 
However, it is unlikely that the stream experiences a drag force, since unless it is plowing through the virial shock for the first time (a highly unlikely scenario), there is no hot gas in front of it. Furthermore, any deceleration of the leading edge of the stream will cause a pile-up of stream material, and thus momentum transfer in the stream's forward direction, an effect that is absent with clouds. Furthermore, for spherical cold clouds, the braking due to momentum conservation only occurs after a free-fall time, following an initial phase where the cloud is free-falling. This timescale is expected to be comparable to the halo-crossing time of the stream, suggesting that the streams should be close to free fall and far from the terminal velocity. However, the differences in the expected evolution of streams versus clouds in an external potential are not yet understood.

\smallskip
In this paper, we use numerical simulations to study the evolution of cold streams in the hot CGM subject to hydrodynamics, radiative cooling, and the external gravitational potential of the host halo. We aim to test and refine the analytic models of M20b for the evolution of cold streams flowing into the dark matter halo potential well. We improve the existing analytic model by relaxing assumptions on the density profile and the resulting shape of the stream and the density profile of the CGM. For the latter, we assume a hot CGM in hydrostatic equilibrium in a Navarro-Frenk-White (NFW) profile \citep{NFW} rather than a simple power law as assumed in M20b. We also explore the effects of the self-shielding of dense gas from the UV background on the emitted radiation, in relation to the matching observed Ly$\alpha$ blobs. Finally, we wish to understand potential similarities and differences between the evolution of cylindrical streams and spherical clouds. 

\smallskip
As a specific application, we relate the issue of stream evolution to the following long-standing puzzle. Massive galaxies of $\gtrsim 10^{11}\msun$ in baryons at $z\sim 2$ reside in halos with virial masses $M_{\rm vir}\gtrsim 10^{12}\msun$. The CGM of these galaxies is expected to contain a hot gas with $T\gtrsim 10^6\K$ in approximate hydrostatic equilibrium. However, the star formation rates (SFRs) observed in these galaxies of $\gtrsim 100\sy$ are significantly larger than expected based on the cosmological accretion rate onto the halo, assuming a maximum efficiency of 1 for transferring gas from the virial radius to the galaxy. Their high star formation rates cannot be explained by mergers \citep{Dekel09}, and are usually much higher than what is predicted from various hydrodynamic simulations \citep{Mitchell2014,Leja2015,Dave2019} as well as analytic bathtub or regulator-type models \citep{Dave2012, Lilly2013, Dekel2014}, where this manifests as the specific SFR (sSFR) being larger than the specific accretion rate (sAR) onto the halo. While new measurements of the SFR at these redshifts yield a lower sSFR, more in line with the predictions of some models and simulations, an offset of a factor $\sim$(2-3) remains \citep{Leja2022}. We wish to quantitatively examine whether the entrainment of hot CGM gas onto the cold streams as they travel from the virial radius to the central galaxy can boost the accretion rate of cold gas onto galaxies enough to explain this discrepancy.

\smallskip
The rest of this paper is organised as follows.
In \cref{sec:theory}, we review the current theoretical understanding of KHI in the presence of radiative cooling and the halo potential. In \cref{sec:sim}, we describe a suite of numerical simulations designed to study stream evolution in the halo potential. In \cref{sec:res}, we present the results of our numerical analysis and compare them with our analytical predictions. In \cref{sec:galaxy_formation}, we discuss the implications of our results for the cold gas accretion rate and star formation rate of galaxies across halo masses and redshifts. In \cref{sec:disc}, we discuss potential caveats to our analysis and outline future work. Finally, we summarise our main conclusions in \cref{sec:conc}.

\section{Theoretical Overview} 
\label{sec:theory}

\subsection{KH Instability}
\label{sec:khi}
KHI arises from shearing motion at the interface between two fluids, leading to the formation of a turbulent mixing layer between the two (sometimes referred to as a shearing layer). This efficiently mixes the two fluids and smooths out the initial contact discontinuity. We focus here on the recent results of \citetalias{M20a}, who analysed the non-linear evolution of KHI in the presence of radiative cooling in a cool, dense cylinder streaming through a hot static background, in three dimensions. 
The system is characterised by three dimensionless parameters: the Mach number of the stream velocity with respect to the background sound speed, $\Mb\equiv \Vs/\cb$; the density contrast between the stream and the background, $\delta\equiv \rhos/\rhob$; and the ratio of the cooling timescale in the mixing layer to the stream disruption time $\tau \equiv t_{\rm cool, mix}/t_{\rm dis}$ (these timescales are defined below). 
Below, we briefly outline the model of \citetalias{M20a} and describe our improvements to that model.

\smallskip
The nature of the instability depends primarily on the ratio of the stream velocity to the sum of the two sound speeds (\citetalias{M16,P18,M18b}),
\begin{equation}
\label{eq:Mtot}
M_{\rm tot}=\frac{\Vs}{\cs+\cb}.
\end{equation}
\smallskip
For planar slabs, the instability is dominated by surface modes for $M_{\rm tot}<1$ and body modes for $M_{\rm tot}>1$. However, for cylinders, azimuthal surface modes are still unstable at $M_{\rm tot}>1$ and dominate the instability for all relevant values of the Mach number (\citetalias{M18b,M20a}). These are concentrated at the fluid interface and lead to the growth of a shear layer that expands into both fluids self-similarly through vortex mergers. Thus, a highly turbulent medium develops within the expanding shear layer, efficiently mixing the two fluids. Independent of the initial perturbation spectrum, the width of the shear layer, $h$, evolves in the absense of cooling as 
\begin{equation}
\label{eq:shear_growth}
h=\alpha \Vs t ,
\end{equation}
{\no}where $\alpha$ is a dimensionless growth rate that depends primarily on $M_{\rm tot}$, and is typically in the range $\alpha\sim 0.05-0.25$ 
\citepalias{P18, M18b}. 
While the width of the shear layer differs in the presence of strong cooling, the stream disruption timescale can be defined as the time when the non-radiative shear layer width grows 
to the size of the stream radius, $h=R_s$, \citepalias{M20a}
\begin{equation}
\label{eq:tau_dis}
t_{\rm dis} = \frac{\Rs}{\alpha\Vs}.
\end{equation}

\smallskip
As the shear layer expands into the background, it mixes the hot CGM gas and the cold dense stream gas. The mean density and temperature in the mixing layer are given by 
\begin{align}
\rho_{\rm mix} &= \sqrt{\rho_{\rm s} \rho_{\rm b}},\\
T_{\rm mix} &= \sqrt{T_{\rm s} T_{\rm b}},\label{eq:mix}
\end{align} 
which can be shown by considering the flux of hot and cold gas into the mixing layer \citep{Begelman1990,GronkeOh2018} or alternatively by conservation of mass and energy \citep{Hillier2019}. The mixed gas in the mixing layer is out of thermal equilibrium, and the relevant cooling time scale is given by
\begin{equation}
\label{eq:mix_cool}
t_{\rm cool, mix} = \frac{k_B T_{\rm mix}}{(\gamma-1) n_{\rm mix}\Lambda(T_{\rm mix})},
\end{equation}
where $n_{\rm mix}$ is the number density in the mixing region, and $\Lambda_{\rm mix}$ is the cooling function at the mean temperature and metallicity in the mixing layer.

\subsubsection{Slow Cooling: Stream Disruption}

When $\tau\equiv t_{\rm cool, mix}/t_{\rm dis}>1$, cooling in the mixing layer is slower than the hydrodynamic disruption of the stream. In this case, the stream will fully mix into the background. The mixing layer expands into the background following \equ{shear_growth}, and the density within the stream decreases, as does the cold gas mass fraction, as the stream mixes into the CGM. Furthermore, material initially in the stream begins pushing on CGM material in the mixing layer, distributing the initial stream momentum over more and more mass as the mixing layer expands. This causes the stream to decelerate over time, with the stream velocity given by (\citetalias{M18b})
\begin{equation}
V_{\rm s}(t) = \frac{V_{\rm s,0}}{1+t/t_{\rm dec}},
\end{equation}
{\no}where $V_{\rm s,0}$ is the initial velocity of the stream, and the deceleration time scale is given by
\begin{equation}
\label{eq:tau_dec}
t_{\rm dec} = \dfrac{\left(1+\sqrt{\delta}\right)\left(\sqrt{1+\delta}-1\right)}{\alpha\sqrt{\delta}}\frac{\Rs}{V_{\rm s,0}}.
\end{equation}
{\no}This is the time when the background mass entrained in the shear layer equals the initial stream mass, such that momentum conservation implies that the velocity is half its initial value.

\subsubsection{Fast Cooling: Stream Survival and Growth}

When $\tau<1$, the cooling time is faster than the stream disruption time. In this case, the gas in the mixing layer cools and condenses onto the cold gas stream efficiently, resulting in net entrainment of hot gas onto the cold stream. Similar to the entrainment of hot gas onto cold clouds \citep{GronkeOh20,Ji2019}, the mass entrainment rate onto the stream can be written as 
\begin{equation}
\dot{m} = \rho_{\rm b} \frac{A}{L} v_{\rm mix},
\end{equation}
where $\dot{m}$ is the rate of change of cold stream mass per unit length (hereafter line-mass), $A$ is the effective surface area of the mixing layer, $L$ is the length of the stream, $v_{\rm mix}$ is the characteristic velocity of the flow through the mixing layer onto the stream. Note that the length of the stream is measured arbitrarily and will cancel out as we measure the line-mass of the stream. The velocity in the mixing layer scales as 
\begin{equation}\label{eq:vmix}
v_{\rm mix} \propto c_{\rm s} (t_{\rm cool,s} /\tsc)^{-1/4},
\end{equation}
where $t_{\rm cool,s}$ is the cooling time of the cold stream. In practise, \citetalias{M20a} found that the minimal cooling time occurs at $T \sim 1.5T_{\rm s}$, and the entrainment rate is well described using $t_{\rm cool,1.5T_s}$ as the cooling time in \Cref{eq:vmix}.

Combining these expressions, the line-mass of the cold stream can be approximated as
\begin{equation}\label{eq:mentr}
m(t) = m_0 \left(1+\frac{t}{\tent}\right),
\end{equation}
where $m_0 $ is the initial line-mass, and $\tent$ is the entrainment timescale given by
\begin{equation}
\label{eq:entrainment}
\tent = \frac{\delta}{2} \left(\frac{t_{\rm cool,1.5T_{\rm s}}}{\tsc}\right)^{1/4} \tsc.
\end{equation}
As hot gas is entrained onto the stream, the stream decelerates due to momentum conservation. The mean velocity of the stream is given by
\begin{equation}\label{eq:ventr}
V_{\rm s}(t) = \frac{V_{\rm s,0}}{1+t/\tent},
\end{equation}
where the deceleration timescale is the same as the mass entrainment timescale. \citetalias{M20a} have shown \cref{eq:mentr,eq:ventr} accurately describe the mass and velocity evolution of cold streams in numerical simulations.


\subsection{Analytic Model in Idealised Halos}
\label{sec:halo_potential}
An analytic model for the evolution of cold streams as they interact with hot CGM in a halo gravitational potential was presented in \citetalias{M20b}. The model begins by deriving the expected properties of cold streams in the outer CGM as a function of halo mass and redshift, assuming that they are in thermal equilibrium with the UV background, in pressure equilibrium with a hot CGM at the virial temperature, and that they carry a large fraction of the total baryonic accretion rate onto the halo. For a halo of $\Mv\sim 10^{12}\msun$ at $z\sim 2$, we expect the density of the streams in the outer CGM to be $n_{\rm H,0}\sim (10^{-3}-0.1)\cmc$, the density contrast between the stream and the CGM to be $\delta_0\sim (30-300)$, the stream radius to be $R_{\rm s,0}\sim (0.03-0.3)\Rv$, and the stream velocity to be $V_{\rm s,0}\sim (0.5-1.5)\Vv$, with $\Vv=(G\Mv/\Rv)^{1/2}$ the halo virial velocity. For a $\Mv\sim 10^{12}\msun$ halo at $z\sim 2$, $\Rv\sim 100\kpc$ and $\Vv\sim 200\kms$ \citep[e.g.][]{Dekel13}.

The model then addresses the properties of streams and the potential entrainment of CGM gas onto them as a function of halocentric radius, as the streams flow towards the halo centre. \citetalias{M20b} assume a simple power-law profile for the CGM density with $\rho\propto \rhalo^{-\beta}$, where $\rhalo$ is the radius with respect to the centre of the dark matter halo. The model also assumes an isothermal equation of state for both the CGM and the stream, and assumes that the two are in local pressure equilibrium at each $\rhalo$. Therefore, the density contrast, $\delta$, between the stream and the CGM remains constant throughout the halo, and the hydrogen number density in the stream is 
\begin{equation}
\label{eq:nhalo}
n_{\rm H,s}(\rhalo) = n_{\rm H,0} \left(\frac{\rhalo}{\Rv}\right)^{-\beta}.
\end{equation} 

The radial density profile leads to a radial profile in the stream cross-section, with the stream growing narrower as it flows towards the halo centre. The profile of stream radius as a function of halocentric radius depends as well on the line-mass profile of the stream, which varies due to mixing and entrainment,  
\begin{equation}
\label{eq:rstream_rhalo}
\Rs(\rhalo) = R_{\rm s,0} \left(\frac{\rhalo}{\Rv}\right)^{\beta/2} \left(\frac{m(\rhalo)}{m_0}\right)^{1/2}, 
\end{equation}
where $R_{\rm s,0}$ is the stream radius at $\Rv$ and $m_0$ is the line-mass as the stream enters $\Rv$. 

The profiles of stream density and radius result in an entrainment timescale that varies with $\rhalo$, becoming shorter as the stream penetrates further into the halo. The entrainment time depends on the cooling and sound crossing times via \Cref{eq:entrainment}. The sound crossing time, $\tsc(\rhalo)\equiv\Rs(\rhalo)/c_s$, is 
\begin{equation}
\label{eq:tsc_halo}
\tsc(\rhalo) = \tsc(\Rv) \left(\frac{\rhalo}{\Rv}\right)^{\beta/2} \left(\frac{m(\rhalo)}{m_0}\right)^{1/2}.
\end{equation}
By combining \Cref{eq:mix_cool,eq:nhalo} while assuming constant temperatures and density contrast, the cooling time becomes 
\begin{equation}
\label{eq:tcool_halo}
t_{\rm cool}(\rhalo) = t_{\rm cool}(\Rv) \left(\frac{\rhalo}{\Rv}\right)^{\beta}.
\end{equation}
Combining \Cref{eq:tsc_halo,eq:tcool_halo,eq:entrainment} yields the dependence of the entrainment time on the halo-centric radius is then
\begin{equation}
\label{eq:tent_halo}
    t_{\rm ent}(\rhalo) = t_{\rm ent}(\Rv)\left(\frac{\rhalo}{\Rv}\right)^{5\beta/8} \left(\frac{m(\rhalo)}{m_0}\right)^{3/8}.
\end{equation}

Finally, the model derives equations of motion for the stream within the halo, assuming a radial orbit from $\Rv$ towards the halo centre. The model accounts for inward acceleration due to the gravitational potential of the halo and an effective drag force decelerating the stream due to entrainment. The key assumption is that entrainment and subsequent deceleration occur at each halocentric radius on the local entrainment timescale (\equnp{tent_halo}), following the models derived by \citetalias{M20a} and described in \Cref{sec:khi}, which ignored the halo potential. The two key equations that govern evolution are (see Section 4.3 in \citetalias{M20b})
\be 
\label{eq:dV_dr}
\frac{1}{t_{\rm ent}(r)}\frac{m_0}{m(r)}V(r)+\frac{1}{2}\frac{dV^2}{dr} = -\frac{d\Phi}{dr},
\ee 
\be 
\label{eq:dm_dr}
V(r)\frac{dm}{dr} = \frac{m_0}{t_{\rm ent}(r)},
\ee 
where $m(r)$ and $V(r)$ are the stream line-mass and velocity at radius $r$, $m_0$ is the initial line mass at $\Rv$, and $\Phi(r)$ is the halo gravitational potential at $r$.

Using this model, \citetalias{M20b} were able to predict the stream line-mass and velocity as a function of the halocentric radius, as well as the total Ly$\alpha$ luminosity emitted outside each radius $r$ as a result of the entrainment induced by the stream-CGM interaction. In \Cref{sec:res}, we compare the predictions of this model, updated to account for more realistic equilibrium profiles of the stream and the CGM (see \Cref{sec:hydrostatic}), with numerical simulations.

\subsection{Equilibrium Profiles for Streams in the CGM}
\label{sec:hydrostatic}
As evident from \Cref{sec:halo_potential}, our predictions for stream evolution and entrainment of hot CGM gas depend sensitively on the density and radius of the stream as a function of the halocentric radius. \citetalias{M20b} assumed a constant value of $\beta$ throughout the halo. Their fiducial value of $\beta=2$ corresponds to an isothermal CGM and a conic stream, though they also considered values of $\beta=1$ and $3$, motivated by CGM density profiles estimated from simulations and observations \citep[e.g.][]{Fielding17,Singh18}. 

\begin{figure}
    \centering
    \includegraphics[width=0.49\textwidth]{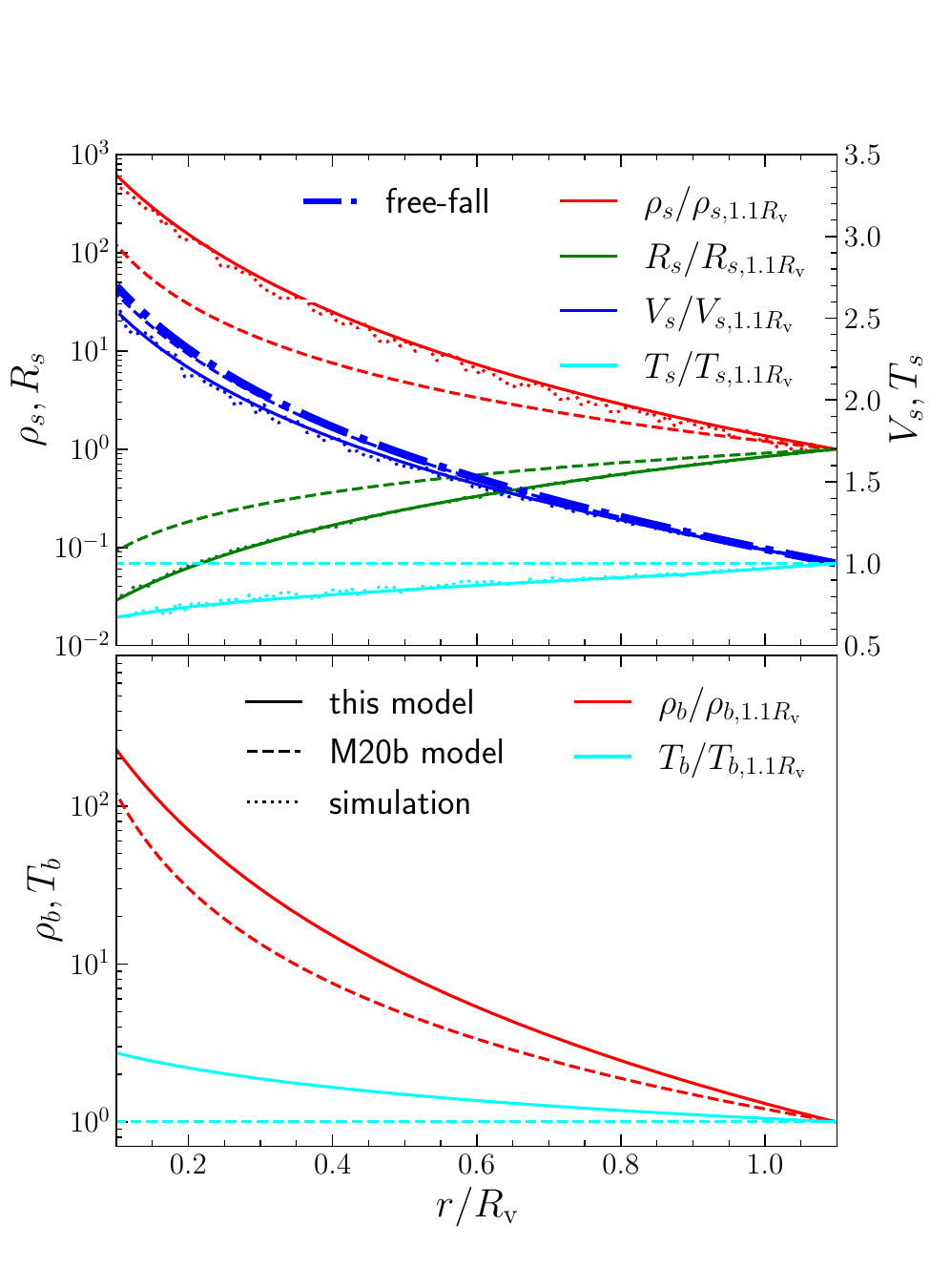}
    \caption{Equilibrium profiles of gas properties in the CGM. The top panel shows the stream properties as a function of distance from the halo centre along the stream normalised by the halo virial radius: the stream density (red) and radius (green) on the left axis, the inflow velocity (blue), and the temperature (cyan) on the right axis. The bottom panel shows the density (red) and temperature (cyan) profiles of the background CGM in which the stream is embedded. Solid lines show our equilibrium profiles based on \equs{stream_pressure}-\equm{stream_radius} for the stream and those based on \citet{KS01} for the CGM, while dashed lines show the simpler model from \citetalias{M20b}. Our new model predicts streams that are denser, narrower, and colder, especially in the inner halo. The velocity profiles in both models are similar and slightly slower than the free-fall velocity, shown by a thick dot-dashed line. The dotted lines show results from simulations without perturbations, after relaxation to an equilibrium state that matches our predictions (the solid lines).}
    \label{fig:model}
\end{figure}

However, here we seek a more realistic equilibrium model for both the streams and the CGM, beginning with a hot CGM in hydrostatic equilibrium within the dark matter halo potential. In other words, we assume that the CGM pressure obeys 
\begin{equation}
\label{eq:Hydrostatic}
\frac{\partial P_{\rm h}({\rhalo})}{\partial \rhalo} = -\rho_{\rm h}({\rhalo})\frac{\partial \Phi}{\partial r} = -\rho_{\rm h}({\rhalo}) \frac{GM(\rhalo)}{\rhalo^2},
\end{equation}
where $M(\rhalo)$ is the enclosed mass as a function of radius that sets the gravitational potential, $P_{\rm h}$ is the pressure of hot gas in CGM of the halo, and $\rho_{\rm h}$ is the hot CGM density. Following \citet{KS01}, we solve \cref{eq:Hydrostatic} assuming that the enclosed mass follows the NFW profile, that the CGM gas is a polytrope $P_{\rm h}\propto \rho_{\rm h}^{\gamma'}$, and that the gas density at large distances follows the dark matter density of the NFW halo. For our chosen halo concentration of $c=10$, the solution corresponds to a polytropic index of $\gamma' = 1.185$ and has a ratio of the halo virial velocity to the CGM sound speed at $\Rv$ of $\Vv/c_{\rm b}\sim 1.45$ \citep{KS01}. Here, $\Vv^2=G\Mv/\Rv$ with $\Mv$ the halo virial mass and $\Rv$ the virial radius, while $c_{\rm b}^2=\gamma P_{\rm b}/\rho_{\rm b}$ with $\gamma=5/3$ the adiabatic index of the gas, and $P_{\rm b}$ and $\rho_{\rm b}$ the pressure and density in the hot CGM at $\Rv$, respectively. Given the CGM density at $\Rv$, which is a parameter that we vary (see \Cref{sec:ini} and \Cref{tab:sim} below), \equ{Hydrostatic} can now be solved for the hot CGM density and pressure profiles. We present the solution in \Cref{app:Hyd_CGM}, and show the resulting equilibrium profiles in the bottom panel of \fig{model}, where the red and cyan lines show the CGM density and temeprature profiles, normalised by their values at $1.1\Rv$, as a function of the halocentric radius normalised by the halo virial radius, $r/\Rv$. The sold lines correspond to our current model, while the dashed lines show the \citetalias{M20b} model with $\beta=2$. In our current model, the CGM is both denser and hotter in the inner halo. 

The inflowing cold stream is assumed to be in pressure equilibrium with the hot CGM at each radius $\rhalo$.  We ignore potential sources of non-thermal pressure, such as magnetic fields or cosmic rays. Therefore, the thermal pressure in the stream and the CGM must be equal at each halo-centric radius, 
\begin{equation}
\label{eq:stream_pressure}
P_{\rm s}({\rhalo})=P_{\rm h}({\rhalo}).
\end{equation}
The gas density in the stream, $\rho_{\rm s}$, is given by 
\be 
\label{eq:stream_density}
\rho_{\rm s}({\rhalo}) = \frac{\mu m_p P_{\rm s}({\rhalo})}{k_BT_{\rm s}({\rhalo})},
\ee 
with $T_{\rm s}\gsim 10^4\K$ the temperature at which the stream is in thermal equilibrium with the assumed UV background \citep[][see \Cref{sec:sim} below]{HM1996}. Neglecting self-shielding effects, $T_s$ is roughly constant, and the stream can be thought of as roughly isothermal. To obtain a more accurate result, we solve for the stream density and temperature profiles as a function of $\rhalo$ iteratively. Given the outer stream density (\Cref{tab:sim}), we compute the equilibrium temperature at that radius and the density profile assuming a constant temperature. We then computed the equilibrium temperature profile for the resulting density profile and repeated the process. The convergence of both profiles to within $<10\%$ is achieved within 3 iterations or less. 

The momentum equation for cold gas inside the stream is given by 
\begin{equation}
\label{eq:stream_velocity}
	\rho_s({\rhalo}) \left(\frac{\partial v_r}{\partial t}+v_r\frac{\partial v_r}{\partial r}\right) = \frac{\partial P_s({\rhalo})}{\partial r} - \frac{GM(\rhalo)\rho_s({\rhalo})}{\rhalo^2}.
\end{equation}
$v_r$ is the radial velocity with respect to the halo as a function of $\rhalo$, and in equilibrium the term $\partial v_r/\partial t = 0$. 
Given the equilibrium pressure and density profiles obtained above, we solve for the stream velocity profile. Neglecting the entrainment of CGM gas onto the stream, which is valid in an equilibrium configuration with no perturbations, mass conservation along the stream implies
\be 
\label{eq:stream_radius}
\rho_s v_r \pi R_{\rm s}^2 = {\rm const},
\ee 
which we can use to obtain the stream radius as a function of the halocentric radius.

We note that in our current model, we ignore the self-gravity of the stream. Thus, all thermodynamic properties are modelled as a function only of halo-centric radius $\rhalo$, with no dependence on the (cylindrical) radial distance from the stream axis or on the polar or azimuthal angles with respect to the halo centre, except for the contact discontinuity at the stream boundary.

The resulting equilibrium profiles for the stream are presented in \Cref{fig:model}. 
We show profiles for the stream density, radius, velocity, and temperature normalised by their values at $1.1\Rv$ (red, green, blue, and cyan solid lines respectively) as a function of the halocentric radius normalised by the halo virial radius, $r/\Rv$, for our fiducial parameters (first row of \Cref{tab:sim}, $V_{\rm s,1.1\Rv}=\Vv$, $n_{\rm H,1.1\Rv}=0.01\cmc$, $\delta_{\rm 1.1\Rv}=100$). We compare these to the corresponding profiles from the \citetalias{M20b} model (dashed lines). Our new equilibrium model predicts streams which are denser, narrower, and colder than in \citetalias{M20b}, though with similar velocity profiles. Unlike the isothermal stream assumed in \citetalias{M20b}, we find that the stream becomes colder at smaller radii, reaching $\sim 0.7T_{\rm s,1.1\Rv}$ at $r\sim 0.1\Rv$, because the higher density corresponds to a lower equilibrium temperature with UV background. The velocity profile is close to the free-fall velocity, but slightly slower due to pressure gradients inside the stream. 
The logarithmic slope of the density profile transitions from $\beta\sim 3.4$ at $1.1\Rv$ to $\beta\sim 1.7$ at $0.1\Rv$. These span the range of constant $\beta$ values assumed in \citetalias{M20b} of $\beta=(1-3)$, though their fiducial value of $\beta=2$ is not representative of the stream in the outer halo.

Using the \citetalias{M20b} formalism with our new equilibrium model, we evaluate the effects of entrainment on stream evolution. These predictions are compared with the results of the simulations in \Cref{sec:sim_res_1}. We note here that the new equilibrium model results in stronger entrainment and, therefore, larger Ly$\alpha$ luminosity than predicted by the original \citetalias{M20b} model. This is primarily due to higher densities and faster cooling rates.

\section{Numerical Methods}
\label{sec:sim}
\smallskip
In this section, we describe the details of our simulation code and setup and our analysis method. We use the Eulerian Adaptive Mesh Refinement code {\sc Ramses} \citep{Teyssier02}, with a piecewise-linear 
reconstruction using the MonCen slope limiter \citep{vanLeer77}, a Harten-Lax-van Leer-Contact (HLLC) approximate Riemann solver \citep{Toro94}, and a multi-grid Poisson solver, to solve the Euler equations for conservation of mass, momentum, and energy in an inviscid fluid:
\begin{align}
\frac{\partial \rho}{\partial t} + \nabla\cdot\left(\rho \vec{v}\right) &= 0, \\
\frac{\partial \left(\rho \vec{v}\right) }{\partial t} + \nabla\cdot\left(\rho \vec{v}\otimes\vec{v}\right) + \nabla p &= -\rho \nabla \Phi, \\
\frac{\partial \left(\rho e\right)}{\partial t} + \nabla\cdot\left[\rho \vec{v}\left(e+\frac{p}{\rho}\right)\right] &= -\rho\vec{v}\cdot\nabla \Phi +\mathcal{H} - \mathcal{C} ,
\end{align}
where $p$ is the thermal pressure and $e$ is the specific total energy with
\begin{equation}
e = \frac{p}{(\gamma - 1) \rho} + \frac{1}{2} v^2. 
\end{equation}
The potential $\Phi$ is given analytically by the NFW dark matter profile in \cref{eq:Hydrostatic,eq:NFW_mass}.
We used the standard {\sc Ramses} cooling module to calculate the cooling term $\mathcal{C}$, which accounts for 
atomic and fine-structure cooling for our assumed metallicity values. The heating term $\mathcal{H}$ is given by photoheating and photoionisation from a $z=2$ UV background given by \citet{HM1996}. For the cases where we included self-shielding of dense gas from the UV background (see \Cref{tab:sim} below), we use the \citet{Rahmati2013} approximation for the photoionisation rate as a function of the gas density, assuming a redshift of $z=2$. We assumed a metallicity of $Z_{\rm s}=0.03Z_{\odot}$ for the stream and $Z_{\rm b}=0.1Z_{\odot}$ for the CGM gas. 

\subsection{Initial \& Boundary Conditions}
\label{sec:ini}

\begin{figure*}
    \centering
    \includegraphics[trim={2.6cm 0.0cm 5.0cm 1cm}, clip,width=0.99\textwidth]{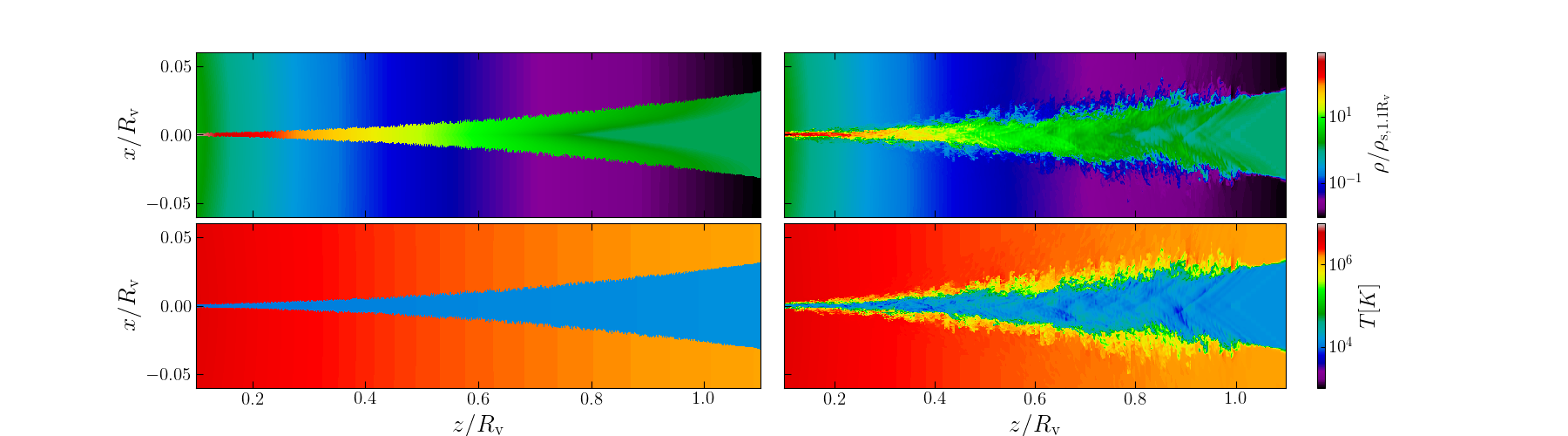}
    \caption{Snapshots of the cold stream embedded within hot CGM from the fiducial simulation (with parameters from first row of \Cref{tab:sim}, $V_{\rm s,1.1\Rv}=\Vv$, $n_{\rm H,1.1\Rv}=0.01\cmc$, $\delta_{\rm 1.1\Rv}=100$). Shown are the 2D maps of the $xz$ plane, the edge-on view of the stream at $y=0$, spanning from $x=[-0.06,0.06]\Rv$ and $z=[0.1,1.1]\Rv$. The left panels show the equilibrium conditions achieved without any perturbations, and the right panels show the stream evolution after running the simulation for 1 box crossing time.  The top panels show the density of the gas, and the bottom panels show the temperature of the gas. The stream flows in from the boundary of the simulation domain at $1.1\Rv$, with an increasing gas density and a decreasing radius of the stream as it approaches the halo centre. }
    \label{fig:map}
\end{figure*}

We set the zero-point of the coordinates at the centre of the dark matter halo. The simulation domain is a cube of side $L=\Rv$. The box sides span the ranges of $z = [0.1\Rv, 1.1\Rv]$, and $x, y = [-0.5\Rv,0.5\Rv]$, such that the centre of the dark matter halo sits outside the simulation box. Placing the inner $0.1\Rv$ outside of the domain allows us to avoid complications associated with this region, which is affected by galactic outflows and gravitational torques from the central galaxy. We assume the stream to be on a radial orbit towards the halo centre with zero angular momentum and place the stream axis along the $z$-axis. We define the radial coordinate with respect to the halo as the spherical radial coordinate, i.e., $\rhalo = \sqrt{x^2+y^2+z^2}$, and the radial coordinate with respect to the stream as the cylindrical radial coordinate, i.e., $\rstream = \sqrt{x^2+y^2}$. 

The stream fluid initially occupies the region $\rstream<\Rso$, where $\Rso$ is the radius of the stream at $1.1\Rv$, while the background (CGM) fluid occupies the rest of the domain. We added an analytic potential corresponding to an NFW dark matter halo with a mass of $10^{12}\Msun$ and a concentration of $c=10$. The virial radius of such a halo at $z=2$ is $\Rv\sim 100\kpc$ \citep[e.g.][]{Dekel13}. The CGM is initially static with $\vec{v_b} = 0$, and its density and temperature are given by hydrostatic equilibrium within this potential, as described in \Cref{sec:hydrostatic} and \Cref{app:Hyd_CGM}. However, we did not initialise the stream properties --- density, temperature, velocity, and radius --- according to the equilibrium profiles shown in \Cref{fig:model}. Rather, the stream is initialised with a constant radius, which we set to $\Rs(r) = R_{\rm s,0}=3\kpc$ in all of our simulations, and a constant velocity $\vec{v_s}(r) = -V_{\rm s,0} \hat{z}$ towards the halo centre, which we vary between $(1.0-1.5)\Vv$ in our simulations (see \Cref{tab:sim}). The initial temperature of the stream is constant and fixed at the thermal equilibrium temperature associated with the density at $1.1\Rv$, $n_{\rm H,0}$, which is another parameter of our simulations (\Cref{tab:sim}). In other words, $T_{s}(r) = T_{\rm s,0}$ where $\mathcal{H}(T_{\rm s,0},n_{\rm H,0}) = \mathcal{C}(T_{\rm s,0},n_{\rm H,0})$. The third parameter of our simulations is the density contrast between the stream and the CGM at $1.1\Rv$, $\delta_0$ (\Cref{tab:sim}), which sets the normalisation for the CGM density profile. The stream density profile is then set such that the stream is in local pressure equilibrium with the CGM at each radius $r$, $P_{\rm s}(r) = P_{\rm h}(r)$.

\smallskip
As we show in \Cref{fig:model} and describe in \Cref{sec:equilibrium} below, the profiles of the stream properties converge to the equilibrium solution within one box-crossing time.
Note that we did not calibrate the total gas mass within our simulation domain based on the universal baryon fraction, though we note that our assumed hydrostatic profiles for the hot component (\equnp{Hydrostatic}) approach the Universal baryon fraction at large $r$ \citep{KS01}. 
The equation of state (EoS) of both fluids is that of an ideal monoatomic gas with an adiabatic index $\gamma=5/3$. 

To prevent the cooling of the hot CGM for long periods of time, we turn off the cooling of the gas with temperatures $T>0.8T_{\rm b}$, with $T_{\rm b}$ the CGM temperature at $1.1\Rv$. Similar methods were employed in \citetalias{M20a} as well as \citet{GronkeOh2018}, though we note that in our case, this makes little difference over one box-crossing time of the stream. 

The stream enters the domain through the $z=1.1\Rv$ boundary and flows along the $z$-axis toward the halo centre at $z=0$. The boundary conditions on this surface are identical to the initial conditions in the stream (as described above) for $\rstream<\Rso$, and identical to the initial conditions in the hot CGM for $\rstream>\Rso$. However, to these we add perturbations to the radial velocity, with respect to the stream axis, 
as described in \cref{sec:methods-pert}. On the other five boundaries of the simulation domain, we use outflow boundary conditions --- also known as zero force boundary conditions. The velocity gradients are set to 0, while the pressure and density gradients, as well as the potential, are taken from the hydrostatic profile computed following \cref{sec:hydrostatic}. 

\subsubsection{Perturbations} 
\label{sec:methods-pert}

\smallskip
The initial conditions of the simulation are unperturbed. Rather, we induce perturbations at each timestep on the inflow boundary condition at $z=1.1\Rv$ from where the stream flows into the domain. As stated above, the stream flows into the simulation box from this boundary with velocity $\vec{v_s} = -V_{\rm s,0} \hat{z}$ at $R<\Rso$, while the background is static, $\vec{v_b}=0$ at $R>\Rso$. We then perturb the radial component of the stream velocity, $v_{\rm R}=v_{\rm x}{\rm cos}(\varphi)+v_{\rm y}{\rm sin}(\varphi)$, with a random realisation of periodic perturbations as in \citetalias{M18b,M20a}. In practice, we perturb the Cartesian components of the velocity, 
\begin{equation}
\label{eq:pertx}
\begin{array}{c}
v_{\rm x}^{\rm pert}(r,\varphi , v_s t) = \sum_{j=1}^{N_{\rm pert}} v_{0,j}~{\rm cos}\left(k_{j}v_s t+m_{j}\varphi + \phi_{j}\right)\\
\\
\times {\rm exp}\left[-\dfrac{(\rstream-\Rso)^2}{2\sigma_{\rm pert}^2}\right]{\rm cos}\left(\varphi\right)
\end{array},
\end{equation}
\begin{equation}
\label{eq:perty}
\begin{array}{c}
v_{\rm y}^{\rm pert}(r,\varphi ,v_s t) = \sum_{j=1}^{N_{\rm pert}} v_{0,j}~{\rm cos}\left(k_{j}v_s t+m_{j}\varphi + \phi_{j}\right)\\
\\
\times {\rm exp}\left[-\dfrac{(\rstream-\Rso)^2}{2\sigma_{\rm pert}^2}\right]{\rm sin}\left(\varphi\right)
\end{array}.
\end{equation}

\smallskip
The velocity perturbations are localised on the stream-background interface, with a penetration depth set by the parameter $\sigma_{\rm pert}$. We set $\sigma_{\rm pert}=\Rs/16$ in all of our simulations, as in \citetalias{M18b,M20a}. 
To comply with periodic boundary conditions, all wavelengths were harmonics of the box length, $k_{j}=2\pi n_{j}$, where $n_{j}$ is an integer corresponding to a wavelength $\lambda_{j}=1/n_{j}$. In each simulation, we include all wavenumbers in the range $n_{j}=2-64$, corresponding to all available wavelengths in the range $R_{\rm s,0}/2 - 16R_{\rm s,0}$. 
Each perturbation mode is also assigned a symmetry mode, represented by the index $m_{j}$ in \Cref{eq:pertx,eq:perty}, and corresponding to the azimuthal modes mentioned in \cref{sec:khi}. As in \citetalias{M18b} and \citetalias{M20a}, we only consider $m=0,1$. For each wavenumber $k_{j}$, we include both a $m=0$ mode and a $m=1$ mode, resulting in $N_{\rm pert}=2\times 63=126$ modes per simulation. Finally, each mode is given a random phase, $\phi_{j} \in [0,2\pi)$. The result is weakly sensitive to changes in random phases, as shown in \citetalias{P18} and \citetalias{M18b}. The amplitude of each mode, $v_{0,j}$, was identical, with the root-mean-square (RMS) amplitude normalised to $0.2\cs$.

\subsubsection{Resolution and Refinement Scheme}
\label{sec:grid_res}
We used a statically refined grid with resolution decreasing away from the stream axis and increasing along the stream as it enters the dark matter halo potential. Near the boundary at $z=1.1\Rv$. The region ${\rm max}(|x|,|y|)<1.5\Rso$ has a cell size of $2^{-11}$ times the size of the box. The resolution then decreases by a factor of 2 every $1.5\Rso$ away from the stream axis in the $\pm x$ and $\pm y$ directions, until it reaches the minimum resolution of $2^{-7}$ times the box size. This is identical to the refinement method employed by \citetalias{M18b} and \citetalias{M20a}. However, to account for the fact that the stream becomes narrower as it nears the halo centre, we add additional levels of refinement at $z<0.6\Rv$ and ${\rm max}(|x|,|y|)<0.75\Rso$ (cell size $2^{-12}$ times the box size) and at $z<0.35\Rv$ and ${\rm max}(|x|,|y|)<0.375\Rso$ (cell size $2^{-13}$ times the box size). 
The initial stream radius is $\Rso = \Rv/32 \sim 3\kpc$, and the best resolution achieved inside the stream is $\sim 46\pc$ at $0.6\Rv<\rhalo<1.1\Rv$ and $\sim 11\pc$ at $0.1\Rv<\rhalo<0.35\Rv$. We find that the results are converged at these resolutions (see \Cref{app:resolution}).

\subsection{Tracing the Two Fluids} 
\label{sec:tracer}
Following \citetalias{M18b} and \citetalias{M20a}, we use a passive scalar field, $\psi(r,\varphi,z,t)$, to track the growth of the shear layer and the mixing of the two fluids. Initially, $\psi=1$ and $0$ in the stream and CGM, respectively. During the simulation, $\psi$ is passively advected with the flow such that the density of the stream fluid in each cell is $\rhos=\psi\rho$, where $\rho$ is the total density in the cell. 

\begin{table}
    \centering
    \begin{tabular}{c|c|c|c|c|c|c|c}
$V_{\rm s,0}/\Vv$ & $\delta_0$ & $n_{\rm H,0}$ & $\tau_{1.1\Rv}$ & $\tau_{0.6\Rv}$ & ${\rm LR}$ & ${\rm HR}$ & s.s. \\
\hline

1.0 & 100 & 0.01  & 0.108 
& 0.082 
& Y & Y & N \\

1.0 & 30 & 0.01   & 0.017 
& 0.009 
& N & N & N \\

1.0 & 300 & 0.01 & 0.650 
& 0.553 
& N & N & N \\

1.5 & 100 & 0.01  & 0.158 
& 0.123 
& N & N & N \\

1.5 & 30 & 0.01   & 0.024 
& 0.013 
& N & N & N \\

1.5 & 300 & 0.01 & 0.961 
& 0.823 & N & N & N \\

1.0 & 100 & 0.1   & 0.006 
& 0.005 
& N & N & N \\

1.0 & 100 & 0.001 & 2.709 
& 1.295 
& N & N & N \\

1.5 & 100 & 0.001 & 3.976 
& 1.942 & N & N & N \\

\hline
1.0 & 100 & 0.01  & 0.108 
& 0.082 
& N & N & Y \\

1.0 & 300 & 0.01 & 0.650 
& 0.553 
& N & N & Y \\

1.0 & 100 & 0.1   & 0.006 
& 0.005 
& N & N & Y \\

1.0 & 100 & 0.001 & 2.709 
& 1.295 
& N & N & Y
\end{tabular}
    \caption{
    Stream parameters in the different simulations at $1.1\Rv$: stream velocity normalized by halo virial velocity ($V_{\rm s,0}/V_{\rm vir}$, roughly 0.69 times the Mach number with respect to the CGM sound speed, \Cref{sec:hydrostatic}), density contrast between the stream and the CGM ($\delta_0$), and the hydrogen number density in the stream ($n_{\rm H,0}$). $\tau$ is the ratio of cooling time to stream disruption time computed based on local stream properties at $1.1\Rv$ and $0.6\Rv$. The final three columns list whether we have a low-resolution (LR) version of the simulation, a high-resolution (HR) version of the simulation, or a version with self-shielding (s.s.) of dense gas from the UV background. }
    \label{tab:sim}
\end{table}

\section{Simulation Results} 
\label{sec:res} 

\smallskip
In this section, we present the results of our simulations. We examine the evolution of cold streams within a hot CGM in hydrostatic equilibrium within an NFW potential, and compare the simulation results with our theoretical predictions.


\subsection{Simulations without Perturbations - Convergence to the Equilibrium Model}
\label{sec:equilibrium}

To test both our numerical setup and the analytic model, we first run simulations without any perturbations. The profiles of stream density, temperature, velocity, and radius after one box-crossing time of the stream for our fiducial parameters (first row of \Cref{tab:sim}) are shown in the top panel of \Cref{fig:model} as a function of the halocentric radius. 
The density, temperature, and velocity of the stream at halocentric radius $\rhalo$ are the $\psi$-weighed average quantity of all gas cells at the same radial bin\footnote{The profiles of volume-weighed quantities of cold gas ($T<T_{1.5T_s}$) are extremely similar}, while the corresponding CGM quantities are weighted by $1-\psi$.
The stream radius is defined as the distance from the stream axis where the azimuthally averaged value of $\psi(\rstream|\rhalo)=0.5$. 

\smallskip
Despite the artificial initial conditions described in \Cref{sec:ini}, the stream profiles have all converged to our analytic equilibrium model described in \cref{sec:hydrostatic} after one box-crossing time.  This highlights both the robustness of our numerical method, as well as the fidelity of our analytical model. We note that when analysing the simulations with perturbations, described in the following sections, we limit our analysis to times after the first box-crossing time, thus avoiding any effects of the artificial initial conditions.

\smallskip
In the left-hand column of \Cref{fig:map}, we show maps of the density (top) and temperature (bottom) for our fiducial simulation with no perturbations after the first box crossing time. The maps represent a thin slice through the mid-plane of the stream. The increase in stream density as it becomes narrower towards the halo centre is apparent, helping the stream to maintain pressure equilibrium with the CGM and a constant mass flow rate. Similarly, the stream temperature decreases slightly towards the halo centre since the higher density corresponds to a lower equilibrium temperature with the UV background. We also see small perturbations in the stream radius which are due to numerical noise. The secondary conic structure inside the stream apparent in the density map is due to the fact that it takes a stream sound crossing time, $t_{\rm s}\sim R_{\rm s}/c_{\rm s}$, for the stream density and pressure to react to the change in CGM density and pressure, while the stream velocity is highly supersonic. Therefore, the stream travels a distance of $l\sim Vt\sim (V/c_{\rm b})\,\delta^{1/2}\,R_{\rm s}$ before the central density increases. For our fiducial values at $1.1\Rv$, ignoring the acceleration of the stream, this corresponds to a distance of $l\sim 0.45\Rv$, consistent with \Cref{fig:map}.


\begin{figure*}
    \centering
    \includegraphics[width=0.33\textwidth]{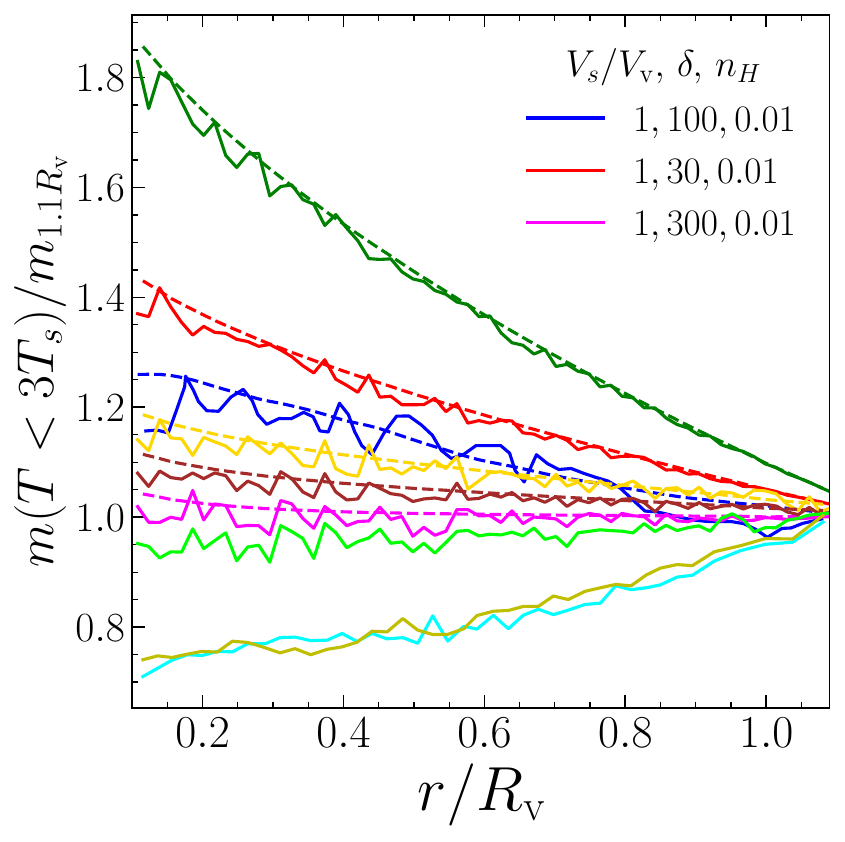}~
	\includegraphics[width=0.33\textwidth]{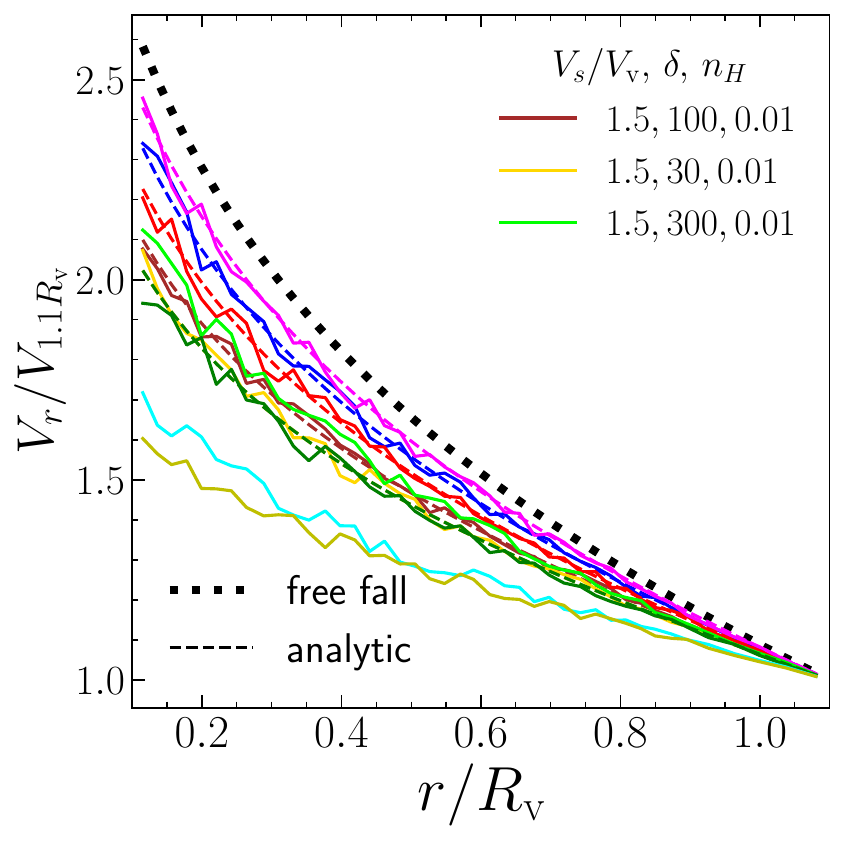}~
	\includegraphics[width=0.33\textwidth]{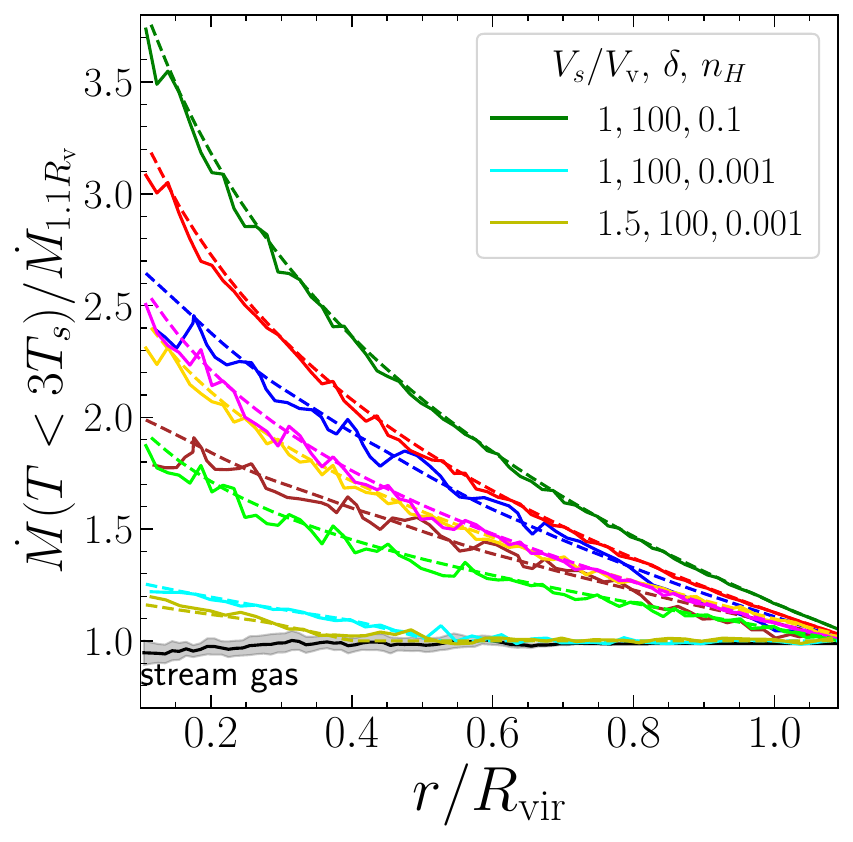}
    \caption{Entrainment of CGM gas onto the cold streams. Shown for the cold gas along the stream are the line-mass (left), the inflow velocity (middle), and mass inflow rate (right), normalised by the value at $\rhalo = 1.1\Rv$. Shown in colours are different cases of stream properties (\Cref{tab:sim}), with the fiducial case for a halo of mass $10^{12}\Msun$ at $z=2$ in blue.
Total cold gas, including the entrained gas (coloured lines), is compared to the original stream gas (black). The simulations (solid) are compared with the analytic predictions (dashed), showing good agreement.
In most cases, as the stream propagates to smaller halo radii, the cold gas is accelerated and the mass inflow rate increases by a factor of 2.5 to 4 due to the entrainment of hot CGM onto the cold streams.  In the case where the stream is very dilute and close to disruption by the KH instability (cyan), the entrainment is weaker. }
    \label{fig:line-mass}
\end{figure*}


\subsection{Entrainment of Hot CGM onto Cold Streams}
\label{sec:sim_res_1}
The right panel of \Cref{fig:map} shows the snapshot of same simulation but with perturbation. We can see the significant changes in the boundary of the stream due to the development of mixing layer, which can rapidly cool onto the stream. \Cref{fig:line-mass} explores the consequences of entrainment and mixing of hot CGM gas onto the cold stream as it flows towards the halo centre. We show radial profiles of the cold gas line-mass (left), the mean inflow velocity of the cold gas (centre), and the cold gas mass accretion rate (right) as a function of halocentric radius. Here and elsewhere, we refer to ``cold'' gas as gas with a temperature less than 3 times the temperature of the stream at $1.1\Rv$, $T<3T_{\rm s}$. In each panel, the stream properties are normalised by their values at $1.1\Rv$, where the stream enters the simulation domain, and the halocentric radius is normalised by $\Rv$. Different coloured lines represent the different simulations listed in \cref{tab:sim}. 
Solid lines show results from simulations with perturbations (\Cref{sec:methods-pert}), and dashed lines show the predictions of our analytic model described in \Cref{sec:theory}. Note that for the line-mass and velocity profiles, we only present analytic predictions for the cases where the stream is in fast cooling regime initially $\tau_{\rm 1.1\Rv}<1$, so the stream line mass is predicted to increase due to entrainment. 
In all cases, the simulation results match the analytic predictions.

We begin by examining the line-mass profiles in the left panel. For most of our simulations with streams in initially fast cooling regime $\tau_{1.1\Rv}<1$ (see \Cref{tab:sim}), the line-mass increases towards smaller radii due to cooling and entrainment of hot CGM gas through the mixing layer onto the cold stream as the latter flows towards the central galaxy. The entrainment rate increases with decreasing $\tau_{1.1\Rv}$. Since we keep both the initial stream radius and the gas metallicity fixed, smaller $\tau_{1.1\Rv}$ values, and thus higher entrainment rates correspond to higher stream density (higher $n_{\rm H}$), higher CGM density (lower $\delta$), or slower streams (lower $V_{\rm s}/\Vv$). Overall, for the cases with $\tau_{\rm 1.1\Rv}<1$, the cold gas mass increases by a few tens of percent as the stream flows from $1.1\Rv$ to $0.1\Rv$, up to $\sim 80\%$ for the range of parameters studied. The three simulations where the cold gas mass decreases between $\Rv$ and $0.1\Rv$, $(V_{\rm s}/\Vv,\delta,n_{\rm H})=(1.5,300,0.01),\,(1.0,100,0.001)$ and $(1.5,100,0.001)$, correspond to the cases where $\tau_{\rm 1.1\Rv}\gtrsim1$. Thus, they are initially in the \textit{slow-cooling} or \textit{disruption} regime, where cooling in the mixing layer is slow compared to the hydrodynamic disruption time of the stream (\citetalias{M20a}). In this regime, the cold gas mass decreases as the stream mixes into the hot CGM rather than the hot CGM being entrained onto the cold stream. However, since the value of $\tau$ decreases towards smaller radii as the stream becomes denser, the decrease in the cold gas mass is not as large as expected from the values at $1.1\Rv$ alone, and the cold gas mass never decreases by more than $\sim 20\%$. Thus, even in these cases, the streams survive down to $0.1\Rv$. 

The velocity profiles in the middle panel show that the cold streams attain net positive acceleration in all of our simulations, though always less than free-fall (shown by the thick dotted black line) due to the entrainment-induced drag force operating on the stream. Among the simulations with $\tau_{1.1\Rv}<1$, the inflow velocity is slower with decreasing $\tau$, namely with faster entrainment. However, the slowest streams are those for which $\tau_{1.1\Rv}>1$ such that the cold streams mix into the hot CGM. However, we note that for these cases, whatever unmixed core of the initial stream remains tends to flow in faster (\citetalias{M18b}). Overall, for the range of parameters studied, the stream velocity at $0.1\Rv$ is $\sim (1.6-2.4)$ the velocity at $1,1\Rv$, or $\sim (60-90)\%$ of the free-fall velocity. 

We note that this is unlike the result found in \citet{Tan23}, who explored the evolution of cold spherical clouds falling through the CGM. They found that clouds exhibited much larger decelerations, eventually reaching a constant `terminal' velocity in some cases. We elaborate on this comparison in \cref{sec:disc}. 

The right-hand panel shows profiles of the radial mass accretion rate, ${\dot{M}}$. The black solid line and shaded region show the mean and $1-\sigma$ scatter among all our simulations of the mass accretion rate of gas initially in the stream at $1.1\Rv$, traced based on the value of the passive scalar $\Psi$ (\cref{sec:tracer}). This remains roughly unity at all radii, showing that the stream gas inflow rate is constant. Only a very small fraction stalls in the CGM due to mixing, leading to an inflow rate of the original gas at $0.1\Rv$ slightly lower than the initial inflow rate. The coloured lines show the mass inflow rate of the cold gas. This is the product of the line-mass and velocity, ${\dot{M}}=\rho AV_{\rm r}=mV_{\rm r}$, where $A$ is the effective cross-section of the cold gas. The cold gas mass accretion rate increases towards smaller radii in all simulations, even those with $\tau_{1.1\Rv}>1$ where the cold gas line-mass decreases towards smaller radii. In most cases, including our fiducial parameters (blue line, $(V_{\rm s}/\Vv,\delta,n_{\rm H})=(1.0,100,0.01)$), the cold mass inflow rate increases by a factor of $\sim (2.5-3.5)$ from $1.1\Rv$ to $0.1\Rv$. This increase is entirely a consequence of the entrainment of initially hot CGM gas onto the stream, as indicated by the fact that the mass inflow rate of gas initially in the stream remains constant. The enhancement of the cold gas inflow rate increases with decreasing $\tau$, similar to the line mass, although opposite to the stream velocity. Our analytic model reproduces these results very well. For the two cases with $\tau_{1.1\Rv}>1$, we note that $\tau$ decreases
towards the halo centre, as both the stream and the CGM become denser (see \Cref{tab:sim}). Thus, while the initial condition at $1.1\Rv$ does not allow the entrainment of hot CGM, the stream may be able to entrain in the inner region. Empirically, we can match the simulation result of ${\dot{M}}$ by assuming that entrainment begins at the radius where $\tau\le 1.2$, and applying our model from this point. This occurs at $\sim 0.5\Rv$ for and $(V_{\rm s}/\Vv,\delta,n_{\rm H})=(1.0,100,0.001)$ and $\sim 0.35\Rv$ for and $(V_{\rm s}/\Vv,\delta,n_{\rm H})=(1.5,100,0.001)$. These predictions are shown in the right-hand panel of \Cref{fig:line-mass}.

\begin{figure}
    \centering
    \includegraphics[width=0.49\textwidth]{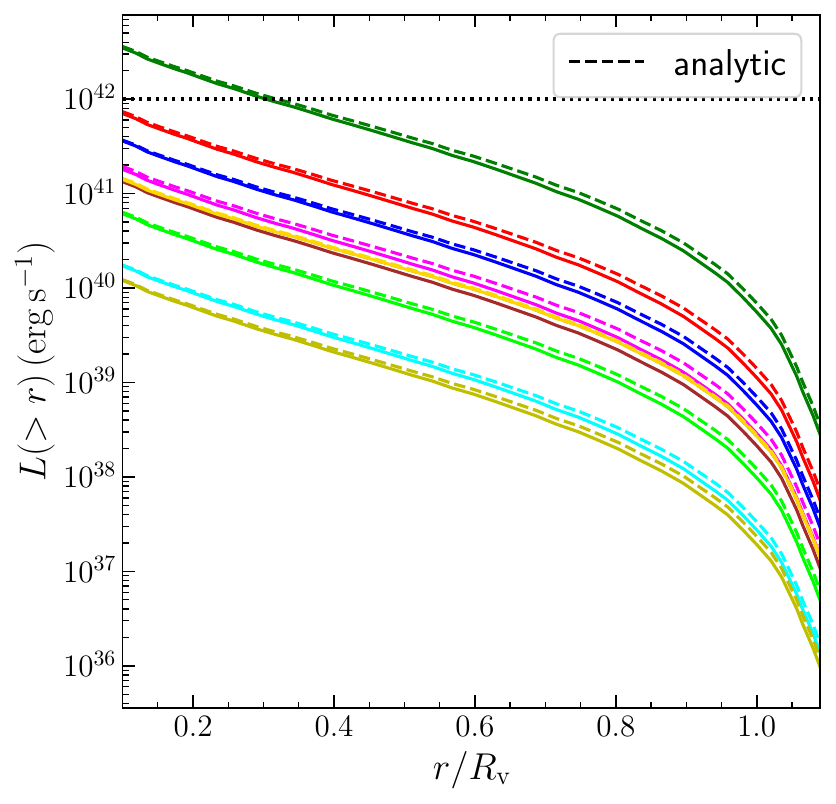}
    \caption{ Lyman-$\alpha$ luminosity profile. Solid lines show the net radiative cooling computed in the simulations, integrated from $r$ to $1.1\Rv$, after subtracting the radiative heating due to the UV background. Dashed lines show our analytic model for the dissipation of mechanical energy and thermal enthalpy due to the entrainment of hot CGM gas onto the cold stream. The two match each other remarkably well. Colours differentiate the different simulations, following the same legend as \Cref{fig:line-mass}. In several simulated cases, the stream-CGM interaction leads to luminosities of $\sim 10^{42}{\rm erg\,s^{-1}}$, consistent with typical observed Ly$\alpha$ blobs. }
    \label{fig:Ldiss}
\end{figure}

\Cref{fig:Ldiss} examines the energy dissipation and cooling emission resulting from the stream-CGM interaction. The dashed lines show our model predictions, based on \citetalias{M20b}, which estimate the total mechanical energy and thermal enthalpy dissipated from $1.1\Rv$ to $r$, for each radius $r$. The solid lines show the cumulative energy radiated from $1.1\Rv$ to $r$, accounting for net cooling minus heating, i.e., excluding the energy input from the UV background, which is subsequently radiated away (so-called UV fluorescence). This is evaluated for a given density, temperature, and metallicity in each simulation cell 
based on the cooling tables and then integrated over all the gas cells outside $r$. The model predictions match the simulation results extremely well, suggesting that all the mechanical and thermal energy dissipated by the stream-CGM interaction is subsequently radiated away. As described in \citetalias{M20a} and \citetalias{M20b}, this is a good proxy for the total Ly$\alpha$ emission from all streams in the halo, because for a given stream, $\sim (30-50)\%$ of the emission is expected to be emitted in Ly$\alpha$, while a given halo is expected to have $\sim (2-3)$ prominent streams. Note that while the difference is very small, the predicted dissipated energy is slightly greater than the measured net cooling emission, because a small fraction of the energy goes to driving turbulence and heating within the stream (\citetalias{M20a}). 

\smallskip
The total luminosity outside of $0.1\Rv$ for streams initially in the fast cooling regime ($\tau_{\rm 1.1\Rv}<1$) is $\sim (0.2-3)\times 10^{42}{\rm erg\,s^{-1}}$, for halos of mass $\Mv\sim 10^{12}\msun$ at $z\sim 2$. Based on \citetalias{M20b}, we expect this luminosity to scale roughly as $L\propto \Mv^{7/6}(1+z)^{2.5}$, yielding a luminosity of $L\sim (1.5-22)\times 10^{42}{\rm erg\,s^{-1}}$ for halos of mass $\Mv\sim 10^{12.5}$ at $z\sim 3$. This is close to the observed luminosity of typical Ly$\alpha$ blobs, $L_{\rm Ly\alpha}\sim (10^{42}-10^{43}) {\rm erg} {\rm s}^{-1}$ \citep{Steidel00,Steidel04,Steidel10,Matsuda06,Matsuda11,Cantalupo14,Borisova16,Leclercq17,Arrigoni18}.
Several of these observed blobs have a filamentary morphology, and it had long been speculated that cold streams may be powering them, provided that they could convert some fraction of the gravitational energy gained by flowing down the potential well of the dark matter halo into Ly$\alpha$ emission \citep{Dijkstra09,Goerdt10,Ceverino15c}. However, these previous works did not specify how the gravitational energy would be converted into radiation. Based on \citetalias{M20b}, we here confirm that the interaction of cold streams with the hot CGM, and in particular the entrainment of the CGM gas through a turbulent mixing layer, can power the emission. 
We note that due to the overall higher stream densities in our model compared to \citetalias{M20b} (\Cref{fig:model}), we predict emission values that are generally a factor of $\sim (2-3)$ larger. 


\subsection{Effect of Self-shielding} \label{sec:sheilding}

\begin{figure*}
    \begin{subfigure}{0.245\textwidth}
    \centering
    \includegraphics[width=1\linewidth]{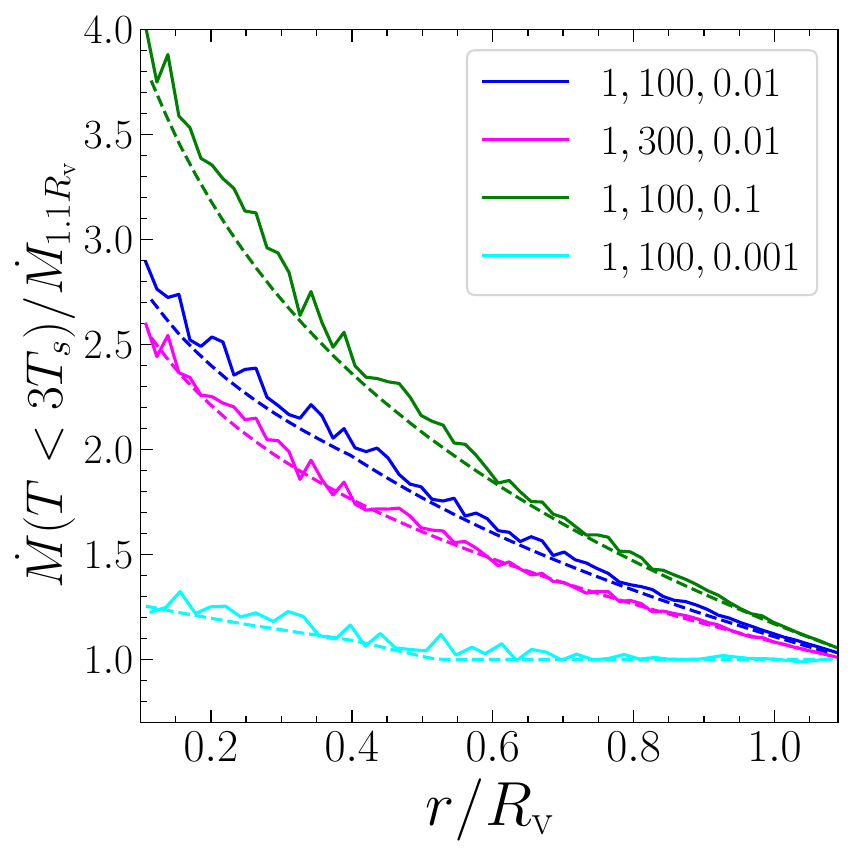}
    \caption{}
    \label{fig:selfshielda} 
    \end{subfigure}
    \begin{subfigure}{0.245\textwidth}
    \centering
    \includegraphics[width=1\linewidth]{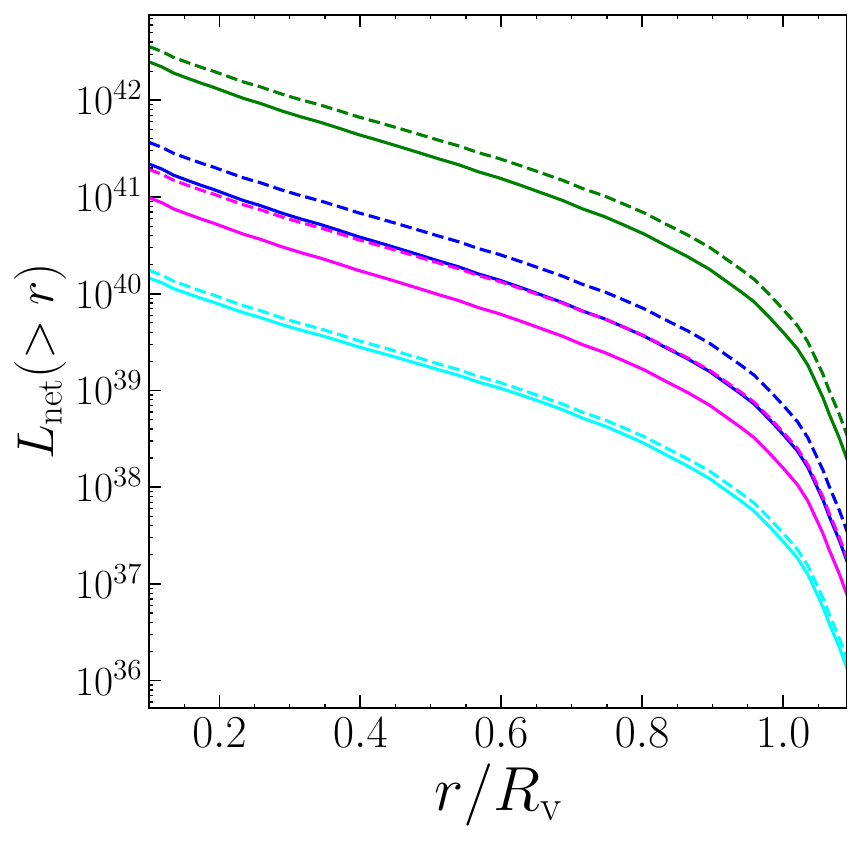}
    \caption{}
    \label{fig:selfshieldb} 
    \end{subfigure}
    \begin{subfigure}{0.245\textwidth}
    \centering
    \includegraphics[width=1\linewidth]{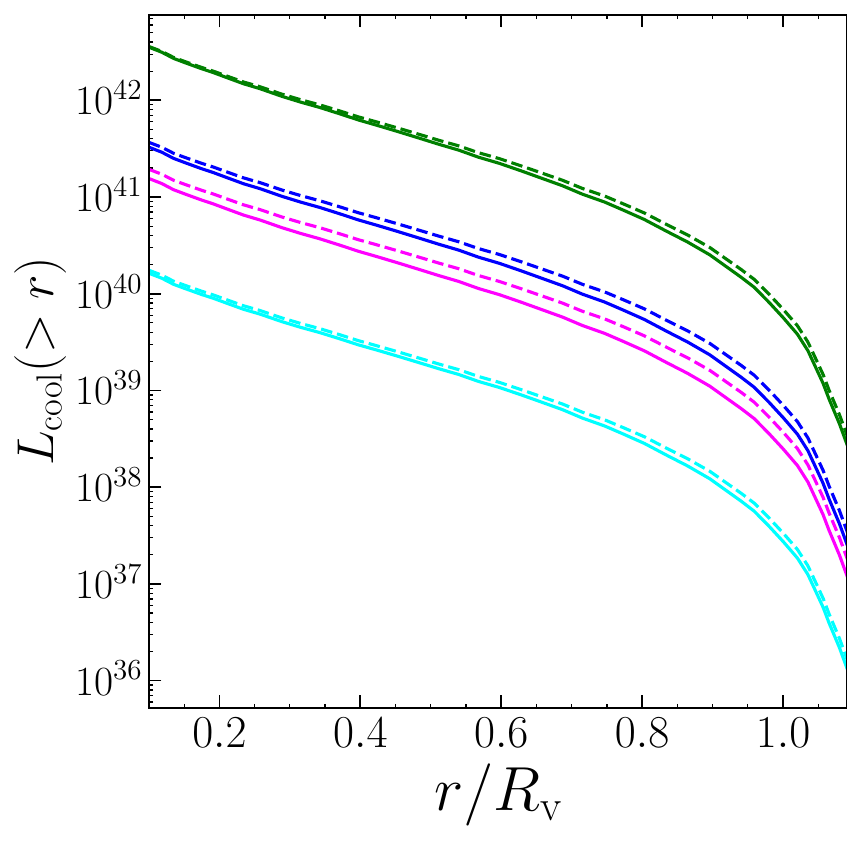}
    \caption{}
    \label{fig:selfshieldc} 
    \end{subfigure}
    \begin{subfigure}{0.245\textwidth}
    \centering
    \includegraphics[width=1\linewidth]{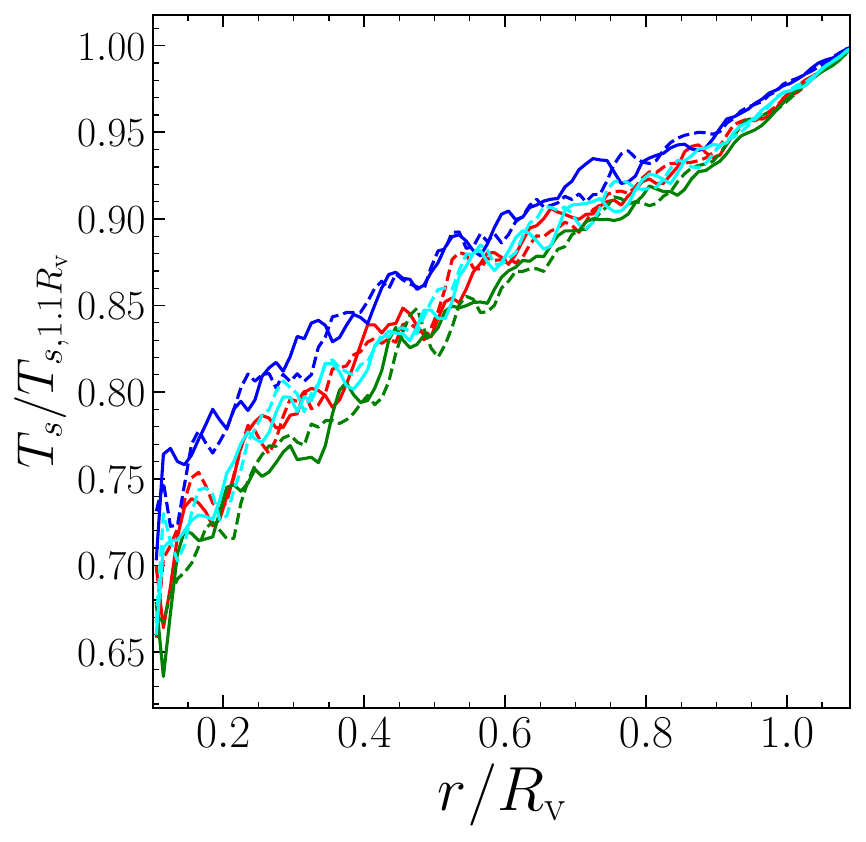}
    \caption{}
    \label{fig:selfshieldd} 
    \end{subfigure}
    \caption{Effect of self-shielding on the stream-CGM interaction. \textit{In panel (a)} we show the mass inflow rate in simulations with self-shielding (solid lines) and our analytic model, which matches simulations without self-shielding (dashed lines). The two agree, showing that self-shielding has not affected the dynamics of entrainment. \textit{In panel (b)} we show the net cooling minus UV heating integrated from $r$ to $1.1\Rv$ in simulations with self-shielding (solid), and our analytic model for the dissipation of mechanical energy and thermal enthalpy induced by the stream-CGM interaction, which is a good match to the net cooling rates in simulations without self-shielding (\Cref{fig:Ldiss}). When self-shielding is included, net cooling rates are a factor of $\sim 2$ lower than energy dissipation rates for stream densities $n_{\rm H,0}\ge 0.01\cmc$. \textit{In panel (c)}, we show the model for energy dissipation (dashed lines, the same as in the middle left panels) alongside the total cooling rates in simulations with self-shielding, including the contribution of UVB. The two agree very well, suggesting that our model for the energy dissipated as a result of the stream-CGM interaction is a good description of the total cooling emission coming from cold streams in a hot CGM. \textit{In panel (d)}, we show the temperature profiles in simulations with self-shielding (solid) and those from simulations without self-shielding (dashed). The two are roughly the same, indicating that self-shielding does not change the overall temperature of the stream. }
    \label{fig:selfshield}
\end{figure*}

The cooling radiation evaluated above and plotted in \Cref{fig:Ldiss} (solid lines) is the net cooling minus UV heating, obtained by subtracting the photoheating due to the UV background from the total cooling rate in each gas cell. As such, it gives a sense of the emission directly produced by the stream-CGM interaction. This is what was modelled analytically by \citetalias{M20b}. However, when comparing it with observations, we must evaluate the total cooling emission, including the UV background fluorescence. In order to properly estimate this, we must consider the self-shielding of dense gas by the UV background. Since most of the emission induced by the stream-CGM interaction is emitted from the intermediate density gas within the mixing layer (\citetalias{M20a}), we might expect self-shielding to have a small effect on this. However, self-shielding will drastically change the total emission coming from the bulk of the stream, and it can also affect the dynamics of the gas during simulation by changing the temperature and pressure of the gas at a given density \citep{FG10}. To examine these effects, we re-ran four of our simulations with self-shielding (\Cref{tab:sim}). 

While the degree to which a gas parcel is self-shielded depends on the total column density, evaluating this for each gas cell during the simulation run-time was computationally prohibitive. Instead, we evaluated the degree of self-shielding of each gas cell during the simulation based on the volume density, following the \citet{Rahmati2013} approximation, widely used in cosmological simulations and based on detailed radiative transfer calculations. We refer the reader to that paper for the full details of the model, but to give a sense, we note here that at $n_{\rm H}=0.001\cmc\,(0.01\cmc)$, the UVB strength is $\sim 95\%$ ($\sim 16\%$) of its unshielded value. 

\Cref{fig:selfshield} compares the evolution of streams with and without self-shielding. \Cref{fig:selfshielda} displays the radial mass flux towards the galaxy, as in the right-hand panel of \Cref{fig:line-mass}. Solid lines represent measurements from simulations with self-shielding, while dashed lines represent the analytic model for the case without self-shielding as in the right-hand panel of \Cref{fig:line-mass}. As shown there, this agrees very well with the measured values from simulations without self-shielding. We see that self-shielding has no noticeable effect on mass flux and, therefore, no noticeable effect on the dynamics of entrainment that governs mass flux (\Cref{sec:sim_res_1}). This supports a scenario in which the entrainment through the mixing layer is not caused by cooling-induced pressure gradients, but rather by turbulent mixing \citep{GronkeOh2018,Fielding20,Tan21}. 

\Cref{fig:selfshieldb} examines the net cooling emission induced by the stream-CGM interaction, as in \Cref{fig:Ldiss}. The dashed lines represent our analytic model for the mechanical energy and thermal enthalpy dissipated through the mixing layer due to the stream-CGM interaction, and are identical to the corresponding dashed lines in \Cref{fig:Ldiss}. As we showed there, this is an excellent approximation for the net cooling emission in simulations without self-shielding. The solid lines in \Cref{fig:selfshieldb} represent the integrated net cooling emission in the simulations with self-shielding, and are thus analogous to the solid lines in \Cref{fig:Ldiss}. When self-shielding is included, the net cooling emission is less than the total dissipated energy by a factor of $\lsim 2$ in all cases where $n_{\rm H}\ge 0.01 \cmc$. This implies that a large fraction of the dissipated energy goes into heating the stream and maintaining its temperature at $T_{\rm s}\sim 10^4{\rm K}$ rather than a few thousand $\K$, which a stream with $Z\sim 0.03 Z_{\odot}$ can, in principle, cool in less than a virial crossing time when ignoring UV heating. Indeed, the temperature profiles of the streams as a function of the halocentric radius, shown in \Cref{fig:selfshieldd}, are very similar with and without self-shielding, gradually decreasing by $\sim 30\%$ from $T_{\rm s}\sim 1.5\times 10^4{\rm K}$ at $1.1\Rv$ to $T_{\rm s}\sim 10^4{\rm K}$ at $0.1\Rv$. 
When self-shielding is not included, the temperature is maintained at $\sim 10^4{\rm K}$ by UV photoheating, while all of the energy dissipated by the stream-CGM interaction is radiated away. However, when accounting for self-shielding, the UV background does not impact streams with densities $n_{\rm H}\gsim 0.01 \cmc$, implying that the energy dissipated by the stream-CGM interaction plays a larger role in maintaining stream temperature in these cases. This argument does not apply to lower-density streams with $n_{\rm H}\sim 10^{-3}\cmc$ (cyan line) where self-shielding is not efficient and the UV background is still capable of heating the stream. In this case, the net cooling measured in the simulation is still a good fit for our prediction of the dissipation rates induced by the stream-CGM interaction. 

In \Cref{fig:selfshieldc}, the dashed lines are the same as in \Cref{fig:selfshieldb}, namely our analytical model for the total energy dissipated as a result of the stream-CGM interaction, which corresponds to the net cooling minus heating rates without self-shielding. The solid lines, however, represent the total cooling rates, including UVB fluorescence, in simulations with self-shielding. This is what will be observed in practice. Consistent with our argument above, we find excellent agreement between the net cooling without self-shielding and the total cooling with self-shielding. The combined energy input from UVB and the dissipation induced by the stream-CGM interaction thus maintain the stream temperature at $T_{\rm s}\sim 10^4{\rm K}$ while generating $\sim (10^{41}-10^{42}){\rm erg\,s^{-1}}$ of radiation, whether or not self-shielding is accounted for. This is consistent with the fact that the same amount of energy is dissipated by the interaction in both cases, as indicated by the fact that the mass accretion rates are the same (\Cref{fig:selfshielda}). All that changes when self-shielding is turned on or off is the relative fraction among the two energy sources (UVB or energy dissipation) that go into maintaining the stream temperature versus generating the radiation. 



\section{Implications for Galaxy Formation}
\label{sec:galaxy_formation}

\smallskip
Our simulations described above confirm the analytic predictions of \citetalias{M20b}. Cold streams survive the journey through the hot CGM towards the central galaxy and grow in mass along the way due to the entrainment of hot CGM gas through a radiative turbulent mixing layer, and the radiation produced by this entrainment-induced-dissipation can match observed Ly$\alpha$ blobs. Furthermore, we find that the cold-gas\footnote{`Cold' here means $T<3T_{\rm s} \lesssim 5\times 10^4{\rm K}$.} mass accretion rate onto the central galaxy is boosted by a factor of $\sim 3$ compared to the accretion rate onto the halo. 
This, in turn, can lead to a similar increase in the galactic SFR above the rate predicted by analytic ``bathub'' or ``equilibrium'' models of galaxy formation, based on the cosmological accretion rate \citep[e.g.][]{Dave2012,Lilly2013,Dekel2014}. Such a boost may be expected in stream-fed galaxies with a volume-filling hot CGM, namely galaxies in halos with $\Mv\gsim 10^{12}\msun$ at $z\gsim 2$.
Previous studies have shown that the observed SFRs in such galaxies are indeed typically higher than those predicted by the cosmological accretion rates by a factor of $2-3$ \citep{Dekel2014}. In what follows, we present an analytical model to assess whether the increase in cold-gas accretion due to entrainment can alleviate this discrepancy.


\subsection{Hot Gas in Halo CGM}
\label{sec:model_halo}
\smallskip
A necessary condition for our model of entrainment-enhanced-accretion is the existence of a hot CGM with $T\sim T_{\rm vir}$ throughout at least most of the range $r\sim (0.1-1.0)\Rv$. This is not the case below a critical halo mass of order $(10^{11.5}-10^{12})M_{\odot}$, where the virial accretion shock around the halos becomes unstable and the CGM is primarily cold and infalling anyway \citep{bd03, Fielding17}. However, cosmological simulations suggest that even below this critical mass scale, the outer CGM may still be hot, even if the inner CGM cools rapidly \citep{Stern2019,Stern2021}. Since hot CGM gas can originate from galactic outflows in addition to virial shock-heating, a better measure of the presence of a hot CGM is the ratio of the local cooling time to free-fall time at each radius $r$, $t_{\rm cool}(r)/t_{\rm ff}(r)$, rather than the global shock-stability criterion \citep{Stern2021}.
Since $t_{\rm cool}\propto n^{-1}$ while $t_{\rm ff}\propto n^{-1/2}$, with $n$ the gas density, this ratio decreases towards smaller radii where the CGM density increases. This results in a picture where the hot CGM develops ``outside-in". The outer CGM virializes first, developing a hot volume-filling component before the inner region \citep{Stern2021}, contrary to the ``inside-out" shock formation scenario of \citet{bd03}. 

To estimate the presence of a hot CGM, we assume the CGM density profile described in \Cref{sec:hydrostatic} and utilised in our simulations. Namely, a CGM in hydrostatic equilibrium within an NFW halo following \citet{KS01}, as a function of the halo mass and redshift. For a given halo mass and redshift, we obtain the normalisation of the density profile (\equnp{CGM_density}) by assuming that the gas density at $\Rv$ is the universal baryon fraction, $f_{\rm b}=0.17$, times the total density at $\Rv$, which follows the NFW profile. We compute the virial radius following the redshift-dependent spherical overdensity criterion of \citet{BN98_vir} and assume a halo concentration following the mass-concentration relation \citep{Diemer2019}.
We then estimate the corresponding profiles of $t_{\rm cool}$ and $t_{\rm ff}$ as a function of $r/\Rv$. The free-fall time at radius $r$ is given by
\be 
\label{eq:tff}
t_{\rm ff}(r) = \left(\frac{3\pi}{32G{\bar{\rho}(r)}}\right)^{1/2},
\ee 
{\no}where ${\bar{\rho}(r)}$ is the mean total density interior to radius $r$, which is given by the NFW profile for total density. The cooling time at radius $r$ is given by 
\be 
\label{eq:tcool}
t_{\rm cool}(r) = \frac{3k_{\rm B}T(r)}{2n(r)\Lambda(T(r))},
\ee 
{\no}where $T(r)$ and $n(r)$ are the CGM temperature and density at radius $r$ respectively, and $\Lambda(T)$ is the cooling curve used in our simulations described in \Cref{sec:sim}. 

\begin{figure}
    \centering
    \includegraphics[width=0.49\textwidth]{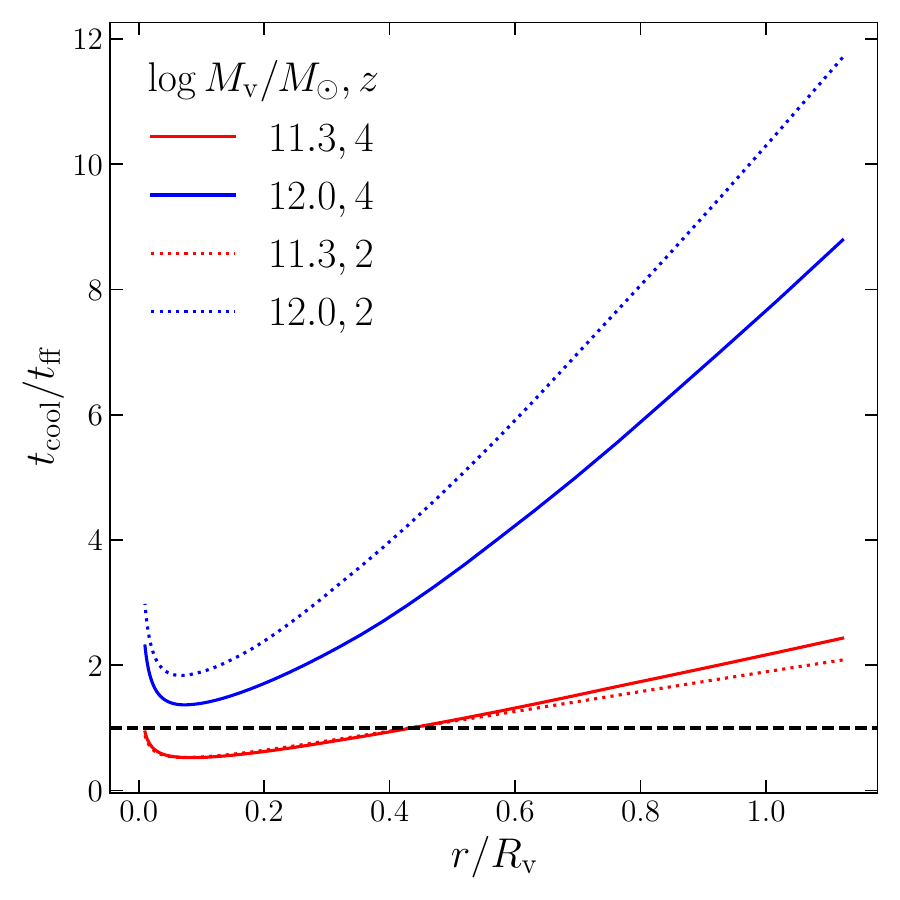}\\
    \includegraphics[width=0.49\textwidth]{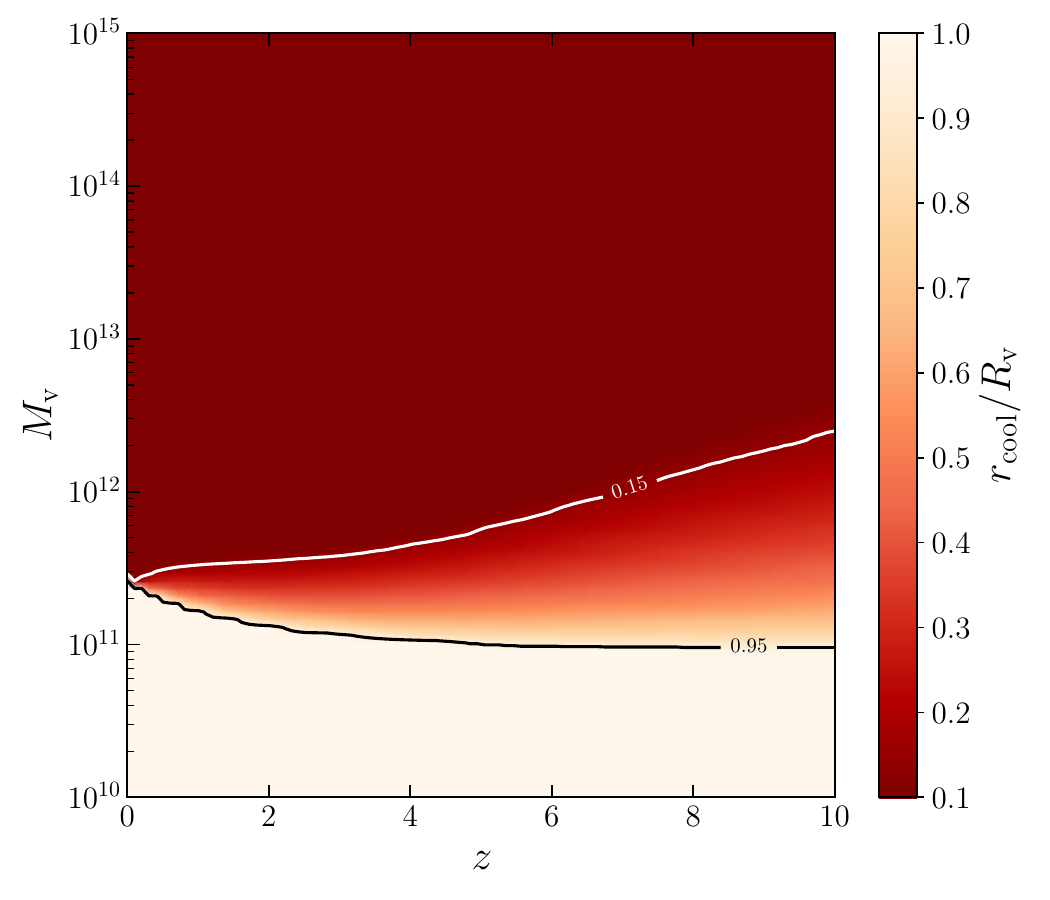}
    \caption{The presence of a hot CGM. The top panel shows the ratio of cooling time to free-fall time as a function of the distance from the halo centre, assuming a hydrostatic CGM profile within an NFW potential, following \Cref{sec:hydrostatic}. For a $10^{12}\Msun$ halo at $z=2-4$ (blue), the cooling time is longer than the free-fall time at all radii, implying that the CGM is hot CGM at all radii. For a $2\times 10^{11}\Msun$ halo at $z=2-4$ (red), $t_{\rm cool}<t_{\rm ff}$ at $r\lsim 0.5\Rv$, implying that the inner half of the CGM is cold while the outer CGM is hot. The bottom panel shows the cooling radius, within which $t_{\rm cool}<t_{\rm ff}$, as a function of the halo mass and redshift. Halos with $\Mv>10^{12}\msun$ and $\Mv<10^{11}\msun$ are all hot and all cold, respectively, at all redshifts. Intermediate-mass halos have a hot outer CGM and a cold inner CGM. The black and white lines mark the contours of $r_{\rm cool}/\Rv=0.95$ and $0.15$ respectively, highlighting the region where the CGM is hot at large radii and cold at small radii.}
    \label{fig:tcooltff_halo}
\end{figure}

The top panel of \Cref{fig:tcooltff_halo} shows the resulting ratio of $t_{\rm cool}/t_{\rm ff}$ as a function of the radius normalised by $\Rv$, for halos of mass $\Mv=10^{12.0}M_{\odot}$ and $\Mv=10^{11.3}M_{\odot}$ at redshifts $z=2$ and $4$. For the $M_{\rm vir}=10^{12.0}M_{\odot}$ halos, $t_{\rm cool}>t_{\rm ff}$ at all radii at both redshifts, implying that the CGM is hot throughout the entire halo. While this is slightly below the \citet{bd03,db06} critical mass scale of $\sim 2\times 10^{12}\msun$, this mass scale was shown to slightly decrease with the inclusion of a central heating source such as galactic feedback \citep{Fielding17}. 
For the $M_{\rm vir}=10^{11.3}M_{\odot}$ halo, which is below the \citet{bd03} critical mass scale, we find that $t_{\rm cool}>t_{\rm ff}$ at $r\gsim 0.5\Rv$ at both redshifts. This implies that the outer CGM is hot while the inner CGM is cold. These results are in line with models of outside-in virialization and hydrodynamical simulations \citep{Stern2019,Stern2021}.
Hereafter, we refer to the outermost radius where $t_{\rm cool}=t_{\rm ff}$ as \textit{the cooling radius}, $r_{\rm cool}$. We assume that CGM is hot at $r_{\rm cool}<r<\Rv$ and cold at $r<r_{\rm cool}$. The bottom panel of \Cref{fig:tcooltff_halo} shows $r_{\rm cool}/\Rv$ as a function of halo mass and redshift. For halos with $\Mv\lsim 10^{11}\msun$, $r_{\rm cool}\sim \Rv$ at all redshifts, which implies that the entire CGM is cold. For halos with $\Mv\gsim 10^{12}\msun$, $r_{\rm cool}\lsim 0.1\Rv$ at all redshifts, which implies that the entire CGM is hot. At intermediate halo masses, the outer CGM is hot, while the inner CGM is cold, with the precise value of $r_{\rm cool}$ varying with both the halo mass and the redshift. We show the contours of $r_{\rm cool}/\Rv=0.95$ and $0.15$, highlighting this transition region.

\subsection{Cold Gas in Filaments}
\label{sec:model_filament}
\smallskip
A second necessary condition for our entrainment model to be valid is the existence of cold streams flowing along the cosmic web filaments that connect to the dark matter halo. Similarly to the CGM in dark matter halos, cosmological simulations suggest that intergalactic filaments can have an outer hot component at the virial temperature of the dark matter filament, which is in virial equilibrium (per-unit-length) within the gravitational potential, surrounding a central cold isothermal core representing the cold stream \citep{Lu2023}. The line mass of a dark matter filament feeding a halo of mass $\Mv=M_{12}10^{12}\msun$ at redshift $(1+z)=5(1+z)_5$ can be evaluated by the cosmological accretion rate onto such a halo \citep{Birnboim16,M18a,Lu2023}, 
\be 
\label{eq:Lambda_general}
\Lambda_{\rm fil} \simeq \rho_{\rm fil}\pi R_{\rm fil}^2 \simeq f_{\rm s}\dot{M}_{\rm v} / V_{\rm s},
\ee 
{\no}where $\rho_{\rm fil}$ and $R_{\rm fil}$ are the characteristic density and radius of the filament, $f_{\rm s}$ is the fraction of the total accretion carried by the given stream, $V_{\rm s}$ is the stream velocity, and $\dot{M}_{\rm v}$ is the total cosmological accretion rate onto the halo given by $\dot{M}_{\rm v}=1572 \Msun \yr^{-1} M_{12}^{1.1} (1+z)^{2.5}_5$ \citep{Fakhouri2010}. In the Einstein-de-Sitter regime, valid at $z>1$, this can be written as \citep{Lu2023}
\begin{equation}
\label{eq:mdot_mfil_sim}
\Lambda_{\rm fil} = 2\times 10^9\Msun \kpc^{-1}\, M_{12}^{0.77} (1+z)^2_5\, f_{\rm s,1/3} \left(V_{\rm s}/\Vv\right)^{-1}, 
\end{equation}
where $f_{\rm s,1/3}=f_{\rm s}/(1/3)$ \citep[see][]{Danovich12}, and $V_{\rm s}/\Vv$ is the ratio of the stream velocity at $\Rv$ to the halo virial velocity, $\Vv$. The hot component is assumed to be at the filament virial temperature, given by \citep{Lu2023}
\begin{equation}
\label{eq:Tvfil}
T_{\rm v,fil} = \frac{\mu m_p G\Lambda_{\rm fil}}{3 k_B}.
\end{equation}
Given the total line-mass of filament in \cref{eq:Lambda_general}, the virial radius of the filament can be written as
\be 
\label{eq:Rvfil}
R_{\rm v,fil} = \left(\frac{\Lambda_{\rm fil} }{\pi \delta_m \rho_m} \right)^{1/2},
\ee
where $\rho_m$ is the mean density of the universe at redshift $z$, and $\delta_m$ is the average overdensity of the filaments estimated to be around $36$ from self-similar cylindrical collapse models (\cref{app:self-similar}; see also \citealp{M18a} for a similar estimate).

The presence of cylindrical accretion shocks around dark matter filaments, analogous to spherical accretion shocks around dark matter halos, was also discussed by \citet{Birnboim16} who devised a shock-stability criterion for filaments similar to the \citet{bd03} shock-stability criterion for halos. Similarly to the \citet{bd03} model for halo shocks, the \citet{Birnboim16} model for filament shocks also assumes that the hot filament forms inside out. However, motivated by the results of \citet{Lu2023} and by our model for the presence of a hot CGM (\Cref{sec:model_halo}), we assume that the hot filament forms outside in, and use the same criterion of $t_{\rm cool}/t_{\rm ff}<1$ to determine the cooling radius within which the filament is cold. 

To evaluate the profiles of $t_{\rm ff}$ and $t_{\rm cool}$, we assume two different density profiles for the filament. The first is an isothermal filament \citep{Lu2023}, where $T_{\rm isothermal}=T_{\rm v,fil}$, and the filament density is given by 
\be 
\label{eq:isothermal_density}
\rho_{\rm isothermal}=\rho_0\left[1+\left(\frac{r}{r_0}\right)^2\right]^{-2},
\ee 
with
\be 
\label{eq:isothermal_scale_radius}
r_0= \left(\frac{2k_{\rm B}T_{\rm isothermal}}{\mu m_{\rm p}\pi G \rho_0}\right)^{1/2},
\ee 
with $\mu= 0.59$ the mean molecular weight of the filament gas, and $m_{\rm p}$ the proton mass. The density normalisation $\rho_0$ is set by the virial radius and total line mass of the filament \cref{eq:Rvfil}.

\begin{figure}
    \centering
    \includegraphics[trim={0.0cm 0.4cm 0.0cm 0.3cm}, clip, width=0.49\textwidth]{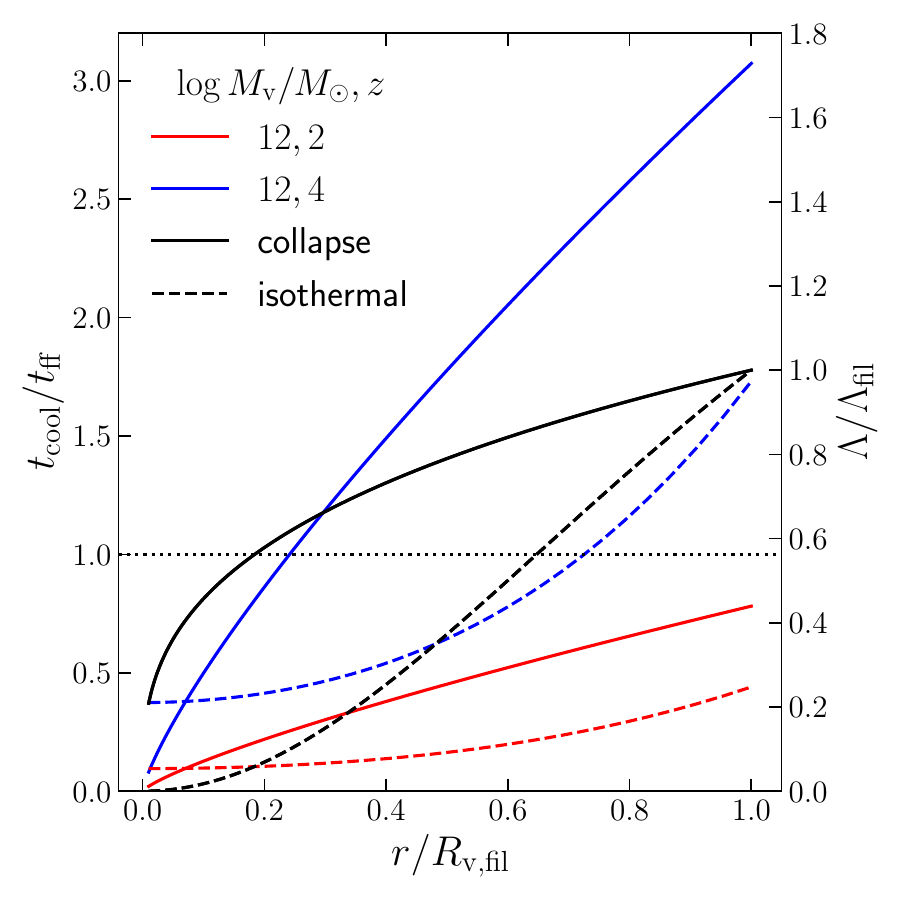}\\
    \includegraphics[trim={0.0cm 0.4cm 0.0cm 0.2cm}, clip, width=0.49\textwidth]{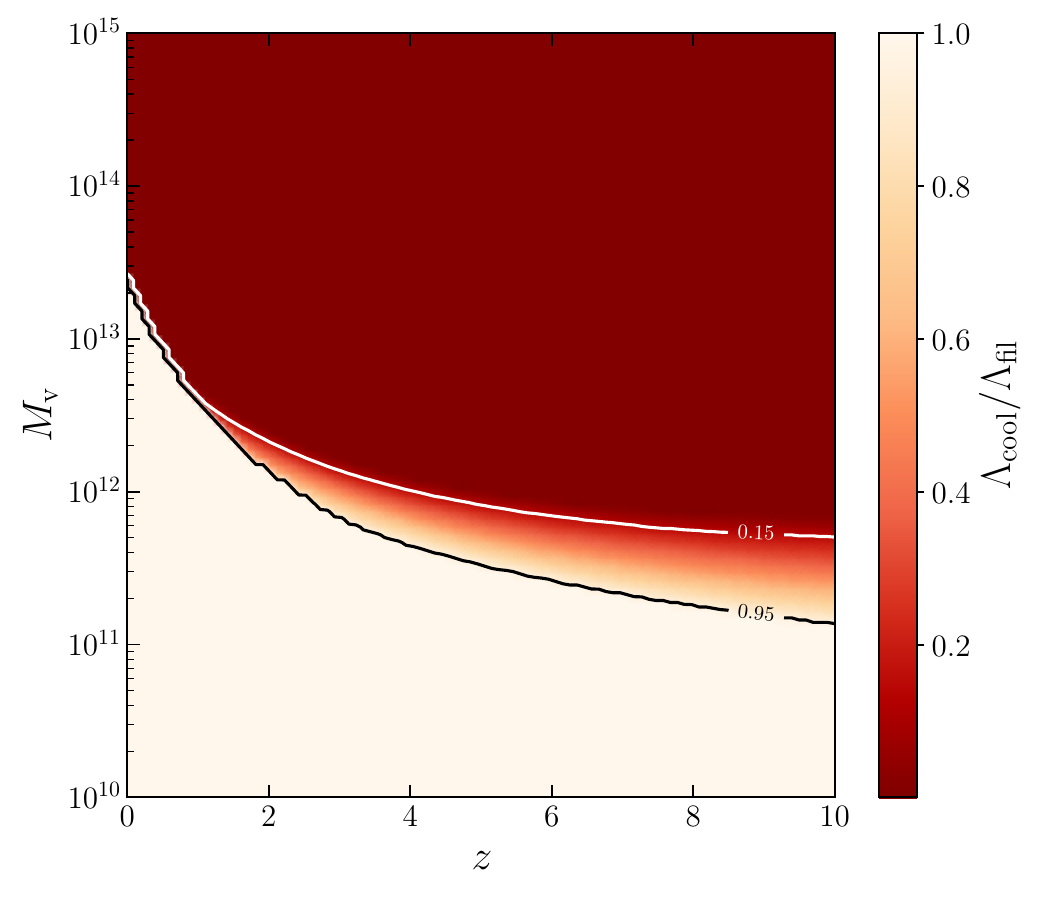}
    \caption{The presence of cold filaments. The top panel shows, as a function of distance to the filament axis normalised by the filament virial radius, $R_{\rm v,fil}$, the ratio of cooling time to free-fall time for a filament feeding a halo of $\Mv=10^{12}\msun$ at $z=2$ (red lines, left $y$ axis) and at $z=4$ (blue lines, left $y$ axis). We also show the normalised enclosed line-mass profile of the filament (black lines, right $y$ axis). The solid (dashed) lines are for our self-similar collapse (isothermal) filament models, respectively. These profiles represent the filament structure outside the halo accretion shock. For filaments at $z=2$, $t_{\rm cool}<t_{\rm ff}$ at all radii, so the filaments are all cold. For filaments at $z=4$, $t_{\rm cool}<t_{\rm ff}$ only at $r\lsim 0.3R_{\rm v,fil}$ or $r\lsim 0.8R_{\rm v,fil}$ for the self-similar filament and the isothermal filament, respectively. Outside these radii, the filament gas is expected to be hot, near the virial temperature (\equnp{Tvfil}). The line-mass profiles of filaments at different redshifts are the same, because both profiles are self-similar. The bottom panel shows the line-mass fraction within the cooling radius, where $t_{\rm cool}/t_{\rm ff}<1$, as a function of the halo mass and redshift. At $z>2$, filaments feeding halos with $\Mv\gsim 10^{12}\Msun$ are mostly hot, while filaments feeding lower mass halos are mostly cold. The transition halo mass increases to $\sim 10^{13}\Msun$ by $z\sim 0$. The black and white lines mark the contours of $\Lambda_{\rm cool}/\Lambda_{\rm fil}=0.95$ and $0.15$, respectively, highlighting the region where the CGM is hot at large radii and cold at small radii.}
    \label{fig:tcooltff_filament}
\end{figure}

The second model we consider for the filament structure is a self-similar model based on the calculation of the self-similar halo gas profile in \citet{bertschinger1985,shi2016b}, but changing the geometry to cylindrical collapse following \citet{FG84}. Details of this model can be found in \Cref{app:self-similar}. 
Both density profiles are normalised using the total line mass and filament radius given in \equs{Lambda_general} and \equm{Rvfil}.

The top panel of \Cref{fig:tcooltff_filament} shows the resulting ratio of $t_{\rm cool}/t_{\rm ff}$ as a function of cross-sectional radius normalised by $R_{\rm v,fil}$ for filaments feeding halos of mass $\Mv=10^{12}M_{\odot}$ at redshifts $z=2$ and $z=4$ (red and blue lines, respectively), assuming the self-similar collapse and isothermal density profiles (solid and dashed lines, respectively). At $z=2$ (red lines), both density profiles yield a filament which is entirely cold, with $t_{\rm cool}<t_{\rm ff}$ at all radii. At $z=4$ (blue lines), the inner filament is cold, while the outer filament is hot outside a cooling radius of $r_{\rm cool}\sim 0.25R_{\rm v,fil}$ and $\sim 0.7R_{\rm v,fil}$ for the self-similar collapse (solid line) and isothermal (dashed line) density profiles, respectively. The solid and dashed black lines show the enclosed line-mass profile of the filament, normalised by its total line mass within $R_{\rm v,fil}$. Since both density profiles have no explicit redshift dependence, these normalised mass profiles are the same at $z=2$ and $z=4$. Comparing the mass profiles to the profiles of $t_{\rm cool}/t_{\rm ff}$ at $z=4$ (blue lines, discussed above), we see that for both density profiles $\sim (65-70)\%$ of the filament line-mass is contained within $r<r_{\rm cool}$. 

The bottom panel of \Cref{fig:tcooltff_filament} shows the fraction of filament mass contained within $r_{\rm cool}$, $\Lambda_{\rm cool}/\Lambda_{fil}$, as a function of halo mass and redshift. We computed this fraction assuming the isothermal filament profile, but the results for the self-similar profile were very similar. At $z\gsim 2$, our model predicts that halos with $\Mv\gsim 10^{12}\msun$ are fed mainly by hot filaments while lower mass halos are fed by mostly cold streams. This threshold halo mass increases to $\sim 10^{13}\Msun$ by $z\sim 0$. This is somewhat contrary to the picture advocated in \citep{db06} where cold streams are more common at $z>2$ than at $z\sim 0$, and where the critical halo mass above which filaments become hot increases with redshift at $z>2$. However, these signify different things and are based on different assumptions. In \citet{db06}, they assume that outside the halo virial shock, the filament is entirely cold and consider whether or not the filament gas will heat up when it penetrates the halo virial shock. On the other hand, here we consider cylindrical accretion shocks surrounding the cosmic web filaments following \citet{Lu2023}, and ask what fraction of the filament gas will be hot versus cold outside the halo virial radius. 

\subsection{Cold Gas Accretion Onto Galaxies across Cosmic Time}
\label{sec:macc_m_z}

\begin{figure}
    \centering
    \includegraphics[trim={0.0cm 0.4cm 0.0cm 0.2cm}, clip, width=0.49\textwidth]{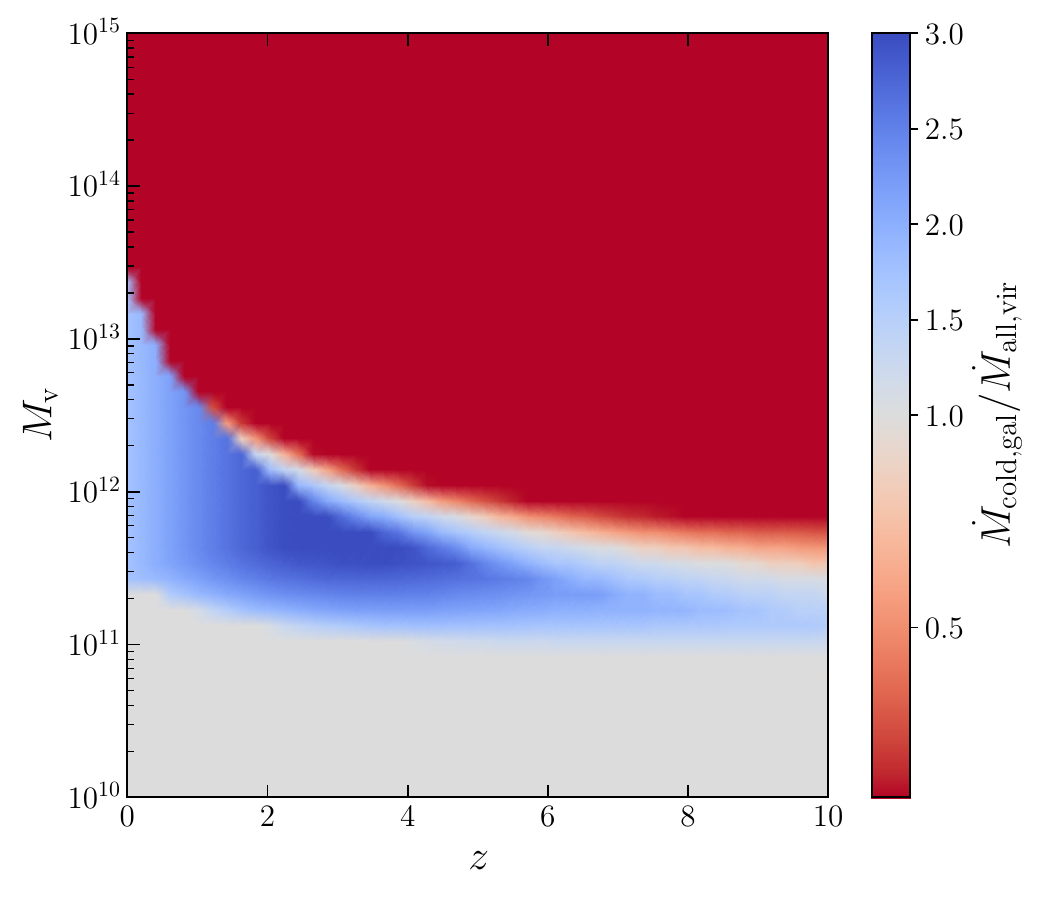}
    \caption{Entrainment as a function of halo mass and redshift. We show the cold gas accretion rate onto the galaxy at $0.1\Rv$, normalised by the total gas accretion rate along the filament at $1.1\Rv$ where it enters the halo. The cold gas fraction in the filaments is evaluated here using the isothermal filament model. In halos with $\Mv\lsim 10^{11}\Msun$ (the grey area), all the gas accreted onto the halo makes it to the galaxy cold. Such halos are in the `cold-flow' regime, where there is no stable virial shock and no hot CGM. In halos above a redshift-dependent threshold mass (red area), ranging from $10^{13}$ at $z\sim 0$ to $10^{11.5}\Msun$ at $z\sim 10$, the cold accretion rate onto the galaxy is lower than the total accretion onto the halo because the cosmic web filaments themselves are mostly hot (see \cref{sec:macc_m_z}). At intermediate masses (blue area), the cold accretion onto the galaxy is larger than the accretion rate onto the halo due to the entrainment of the hot CGM onto the cold streams.}
    \label{fig:macc_m_z}
\end{figure}

\smallskip
We now combine the results of \cref{sec:model_halo} and \cref{sec:model_filament} for the presence (or lack thereof) of cold streams flowing through a hot CGM, with our analytic model presented in \cref{sec:theory} and confirmed with simulations in \cref{sec:res} for the entrainment of hot CGM gas onto the cold stream in such a scenario. Using these tools, we make predictions for the cold gas accretion rate onto the galaxy (assumed to be at $0.1\Rv$) normalised by the total gas accretion rate onto the halo, as a function of halo mass and redshift. We present these results in \Cref{fig:macc_m_z}. The cold gas fraction in the filaments is evaluated here using the isothermal model, though similar trends are seen using the self-similar collapse model. 

At low halo masses, $\Mv\lsim 10^{11}\msun$ with a slightly larger threshold mass at $z\lsim 2$, the cold gas accretion rate onto the galaxy is the same as the gas accretion rate onto the dark matter halo. The boundary of this region is very similar to the boundary for the presence of a hot CGM, namely where $r_{\rm cool}/\Rv\sim 1$ in the bottom panel of \Cref{fig:tcooltff_halo}. Although such halos are fed by cold filaments, they do not have a hot CGM, and therefore entrainment does not play a role. The cold stream may still mix with the ``cold'' gas in the surrounding C/IGM, but since this gas has temperatures similar to the filament gas there is no additional cooling or loss of thermal pressure support. Hence, there should be no noticeable entrainment or increase in the cosmological accretion rate between $\Rv$ and $0.1\Rv$. 

At high halo masses and high redshift, $\Mv\gsim 10^{12}\msun$ at $z\gsim 2$ and $\Mv\gsim 10^{13}\msun$ at $z\sim 0$, we predict that there is little cold gas accreted by the galaxy at all. The boundary of this region is similar to the boundary where $\Lambda_{\rm cool}\sim 0$ in the bottom panel of \Cref{fig:tcooltff_filament}. These halos are fed by hot filaments that do not cool before penetrating the halo virial shock. This would suggest that such haloes are in the so-called ``hot-mode'' accretion regime \citep{Keres05,vdv11}; however, we cannot rule out subsequent cooling of the filament within $\Rv$ due to the higher pressures in the CGM. Note that most of these halos constitute very high $\sigma$-peaks in the cosmic density field, and rarely form at high redshift. At low redshift, these halos comprise galaxy clusters and massive groups whose central galaxies are usually quenched. 

At intermediate halo masses and redshifts, in between these two regions, we find halos with a hot CGM fed by cold streams, providing the right conditions for entrainment. The cold gas accretion rate onto the galaxy in this regime is boosted compared to the gas accretion rate onto the halo by up to a factor of $\sim 3$. This boost occurs over a very narrow range in halo masses at $z\gsim 6$, but at smaller redshifts it spans $\sim 1-2$ dex in halo mass, increasing towards lower $z$. The boost is maximal at $z\sim 1-4$, decreasing towards lower and higher redshifts. 


\subsection{Bathtub Model and Star Formation Rates in Galaxies}\label{sec:bathtub}

\smallskip
To see how the results of the previous section may affect the SFRs of galaxies, we implement them in a bathtub model for galaxy evolution \citep[e.g.][]{Dave2012, Lilly2013, Dekel2014, Mitra2015}. In this model, the gas mass in the interstellar medium (ISM), $M_g$, and stellar mass of the galaxy, $M_*$, are evolved using source and sink terms representing accretion, star-formation, and outflows. Here, we follow the basic framework of the model as presented in \citet{Dekel2014}, and refer the reader to that paper for the full details of all model parameters and assumptions. Here, we provide a brief summary of the key components of the model. The equations governing the evolution of gas and stellar mass in the galaxy are 
\begin{align}
	\dot{M}_g &= (1-f_{sa})\dot{M}_a - (\mu + \eta) \dot{M}_{sf}\label{eq:mgdot_bathtub}\\
	\dot{M}_* &= f_{sa} \dot{M}_a + \mu \dot{M}_{sf}.\label{eq:msdot_bathtub}
\end{align}
$\dot{M}_a$ is the baryonic accretion rate onto the ISM, and $f_{sa}$ is the stellar fraction of the accreted baryons, the rest being gas. $\dot{M}_{sf}$ is the star formation rate and $\mu$ is the fraction of stars that remain locked in long-lived stars or stellar remnants, while the rest are assumed to be instantaneously deposited back into the ISM due to stellar winds and supernova \citep{Tinsley1980}. $\mu$ is estimated to be $0.54$ after $z\sim 2$ \citep{Krumholz2012}, but can be slightly larger at higher redshifts when stellar populations are less evolved. However, even at $z\sim 6$ one expects $\mu\lsim 0.62$, so we adopt a constant value of $\mu=0.54$ at all redshifts, following \citet{Dekel2014}. $\eta$ is the effective mass loading factor, which parameterizes the mass outflow rate from the galaxy due to stellar feedback normalised by the current SFR. Following \citet{Dekel2014}, we can use a lower, or even negative, value of $\eta$ to model gas recycling (outflows that fall back onto the galaxy) in the instantaneous recycling approximation. 

Motivated by theoretical considerations and by the empirical Kennicutt-Schmidt relation \citep{Kennicutt89,Kennicutt98}, the SFR is assumed to be proportional to the current gas mass, 
\be 
\label{eq:SFR_bathtub}
\dot{M}_{sf} = \frac{M_{\rm g}}{t_{\rm SF}},
\ee
where $t_{\rm SF}$ is the \textit{depletion time} \citep[e.g.][]{Genzel08,Dave2012}, namely the time for star-formation to consume the current gas mass, ignoring $\mu$. We assume that this is proportional to the disk dynamical time
\be
\label{eq:tsf}
	t_{SF} = \epsilon^{-1}t_d = \epsilon^{-1}R_d/V_d,
\ee
where $R_d$ and $V_d$ are the characteristic radius and rotation velocity of the disc. This is based on the assumption that giant star-forming clumps in high redshift star-forming galaxies are in the so-called ``Toomre-regime'', with a local free-fall time proportional to the global disk dynamical time\footnote{This is as opposed to the ``GMC-regime'' common in low-$z$ disks where the local free-fall time is decoupled from the global dynamical time \citep{Krumholz.etal.2012}}. \citep{Krumholz.etal.2012,Dekel2014}. The disk dynamical time is assumed to be proportional to the cosmic time \citep{Dekel13,Dekel2014}, 
\be 
\label{eq:td_t}
t_{\rm d} = \nu t,\:\:\:\nu\simeq 0.0071.
\ee 
$\epsilon$ indicates the star formation efficiency per dynamical time and is assumed to be constant over the range of galaxy masses and redshifts considered. This is a free-parameter of the model, with a fiducial value of $\epsilon= 0.02$ \citep[e.g.][]{Krumholz.etal.2012,Dekel2014}.

Bathtub models typically assume that the gas accretion rate onto the galaxy ISM is limited from above by the cosmological gas accretion rate onto the halo and may be lower than this due to some form of ``preventative feedback" \citep{Mitra2015}. However, as shown, this value can be boosted by a factor of up to 3, given the cooling and entrainment of additional CGM gas onto the cold stream. Thus, we assume the accretion rate of cold gas onto the galaxy as 
\begin{equation}
	\label{eq:accretion}
	\frac{\dot{M}_a}{M_a} = s~\frac{\dot{M}_{a,cosmic}}{M_a} \simeq s~0.03\Gyr^{-1} (1+z)^{5/2}.
\end{equation}
The specific cosmological accretion rate, $0.03\Gyr^{-1}(1+z)^{5/2}$, is valid in the Einstein-de Sitter (EdS) regime, with $\Omega_m=1$, which is a good approximation at high redshift, $z>1$. This formula can be derived analytically and is confirmed by numerical simulations \citep{NeisteinDekel08,Dekel13,Dekel2014}. $s$ is the boost factor that increases the accretion rate onto the galaxy compared to the accretion rate onto the halo due to the entrainment of hot CGM gas onto the cold streams (\Cref{fig:macc_m_z}).

\subsubsection{Solution}
\label{sec:sol}
\smallskip
We start by examining the \textit{quasi-steady state} (QSS) solution of the model \citep{Dekel2014}. In this approximation, sometimes called the \textit{equilibrium} solution, the gas mass changes slowly with time, such that $\dot{M_{\rm g}}$ can be neglected in \cref{eq:mgdot_bathtub}. This occurs when the timescale for the gas to reach equilibrium between inflows and star-formation plus outflows is much faster than the timescale on which any of these source/sink terms vary (see the discussion of the conditions for this in \citealp{Dekel2014}). Under this assumption, the gas mass in ISM is given by 
\begin{equation}
	M_g(t) = \frac{(1-f_{sa}) t_{SF} (t)}{\mu + \eta} \dot{M}_a(t), 
\end{equation}
and the specific star formation rate is given by  
\begin{equation}
	{\rm sSFR} = \frac{\dot{M}_{SF}}{M_*} = \frac{(1-f_{sa})}{\mu + f_{sa}\eta} \frac{\dot{M}_a}{M_a}.
\end{equation}

The QSS solution introduces small errors in the gas and stellar mass with respect to the full-time-dependent solution. However, errors in intensive quantities, such as sSFR, which scales as $M_{\rm g}/M_{\rm s}$, as well as the stellar-to-halo mass ratio and the gas fraction, are much smaller \citep{Dekel2014}.
Additionally, the QSS solution introduces a transient error compared to the time-integrated solution at an early time, before the system has reached equilibrium. However, these errors decay over time and are negligible at $z\sim 2$. In practice, we find that the QSS solution is within $\sim 0.1$ dex of the time-integrated solution at redshifts $z<6$, assuming initial conditions at $z=10$ as in \citet{Dekel2014}. This is true even when including the halo mass and redshift-dependent boost factor, $s(M(z),z)$, following \Cref{fig:macc_m_z}. Therefore, when comparing our model to observations in \Cref{sec:obs} we present the results of the QSS solution.


\subsubsection{Stellar Accretion}
\label{sec:stellar_accretion}
\smallskip
The fraction of stellar accretion, $f_{sa}$ is assumed to be 0 in most bathtub models \citep[e.g.,][]{Dave2012, Lilly2013}. Although this is likely the case for lower-mass halos, high-mass galaxies at $z\sim 2$ can have significant ex-situ stellar populations and have a higher fraction of stellar mass accreted \citep{Krumholz2012}. The higher stellar fraction also leads to less sensitivity of the result to other parameters and to suppression of in situ star formation, thus lowering specific star formation rates \citep{Dekel2014}. Here, we argue that $f_{sa}$ is a function of halo mass due to the stellar-mass-halo-mass relation. 

Most stellar accretion comes from accreting other galaxies, following either the merger of two similar mass dark matter halos or the accretion of subhaloes. Thus, the value of $f_{sa}$ can be approximated by the stellar-to-baryonic mass fraction in the dark matter halos merging with our target halo. For halos less massive than $\Mv\sim 10^{12.5}\msun$, this fraction is likely to be maximal for an equal mass major merger, due to the decrease in the stellar-to-halo mass ratio with decreasing halo mass for halos with $\Mv\lsim 10^{12}\msun$ \citep[e.g.][]{Behroozi19}. In this instance, we have $f_{sa}(M)=M_*(M/2)/(f_{\rm b} M/2)$, where $M_*(M)$ is the typical stellar mass for a halo of mass $M$ and $f_{\rm b}\sim 0.17$ is the universal baryon fraction. Less massive mergers will have a lower stellar fraction, while smooth accretion from the cosmic web can be approximated as purely gaseous. Nonetheless, we use this formula for $f_{sa}$ to obtain an upper limit on the contribution of stars to accretion.
For $10^{12}\Msun$ halos, $f_{sa}\sim 0.18$ is roughly independent of the redshift, similar to the maximal value assumed in \citet{Dekel2014}. However, for lower halo masses of $10^{11}\Msun$, we obtain $f_{sa}\lesssim 0.03$, assuming the stellar mass-halo mass relations in \citet{Behroozi19}. 


\subsection{Comparison to Observation}
\label{sec:obs}

\begin{figure}
    \centering
    \includegraphics[trim={0.0cm 0.4cm 0.0cm 0.2cm}, clip, width=0.49\textwidth]{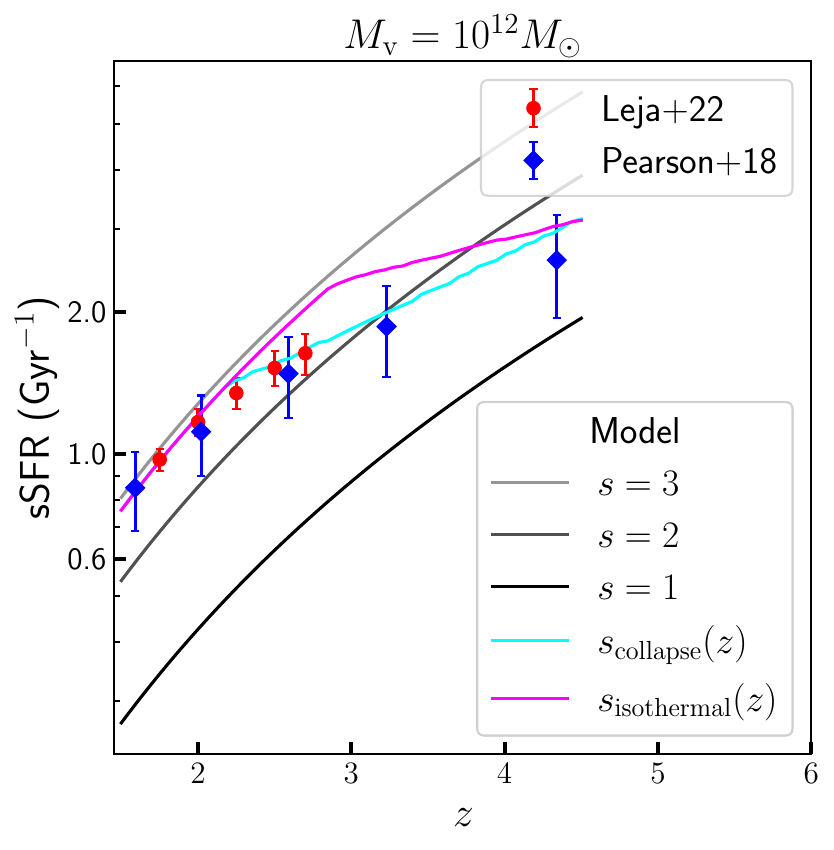}
\caption{Results of our bathtub plus entrainment model compared to observations. We show the specific star formation rate as a function of redshift for halos of mass $\Mv\sim 10^{12}\Msun$. The observational results, shown by coloured symbols, are taken from \citet{Pearson2018} (blue circles) and \citet{Leja2022} (red diamonds). Our model predictions assume a stellar fraction in the accretion of $f_{sa}=0.2$ an outflow mass loading factor of $\eta=3$, and a star-formation efficiency of $\epsilon=0.02$. We show results where accretion is enhanced by an entrainment-induced factor of $s=1$, $2$, or $3$ (from black to grey, respectively). Although $s=3$ can match the data at $z<2.5$, the data at higher-$z$ prefer a smaller boost factor. The coloured lines show the predicted sSFR when the entrainment factor is estimated according to our mass and redshift-dependent model for the cold gas fractions in cosmic web filaments and the CGM (\cref{sec:macc_m_z}). This model reproduces the data because the lower fractions of cold gas in the filaments at $z>2.5$ lower the overall entrainment and the corresponding boost factor. }
    \label{fig:sfr_highmass}
\end{figure}

\subsubsection{Specific Star-Formation Rates of Star-Forming Galaxies}

\smallskip
Here, we wish to examine whether our bathtub plus entrainment model can reproduce the observed star formation rate in star-forming galaxies during cosmic noon, thus offering a potential solution to the apparent paradox of the minimal bathtub model presented in \citet{Dekel2014}. While a detailed Monte-Carlo study of all the different parameter combinations compared against a large compilation of observations is beyond the scope of the current paper and is left for future work, we present results for models with typical parameters (see also \citealp{Dekel2014}) compared to recent observations of sSFR at cosmic noon, as detailed below. 

\begin{figure}
    \centering
    \includegraphics[trim={0.0cm 0.4cm 0.0cm 0.2cm}, clip, width=0.49\textwidth]{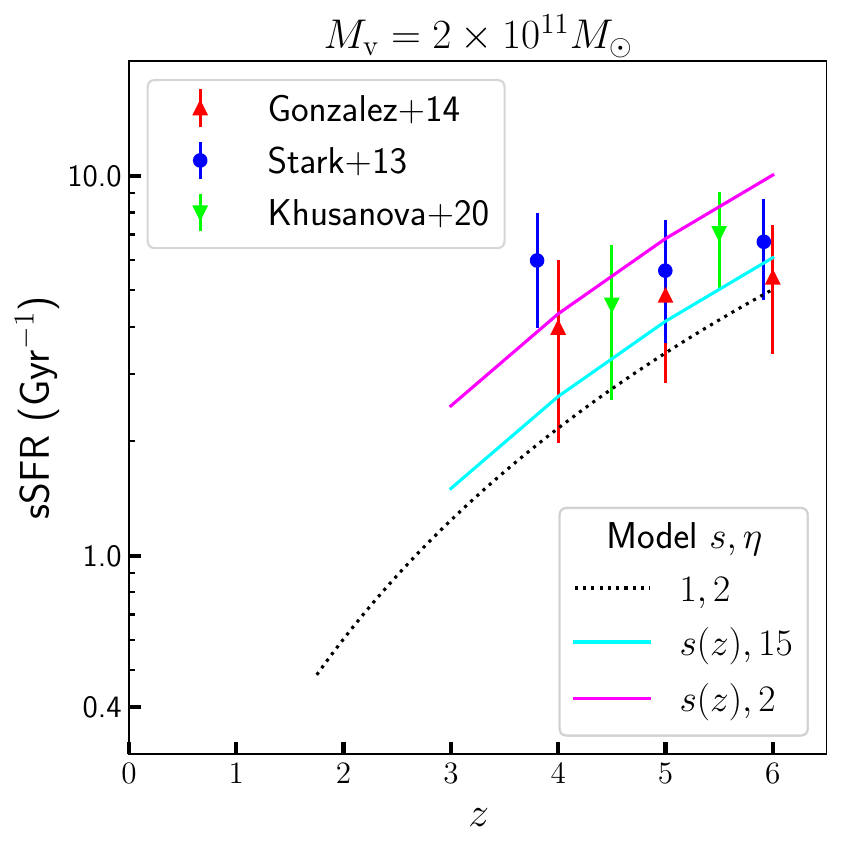}
    \caption{Same as \Cref{fig:sfr_highmass} for halos of $2\times 10^{11} \Msun$ at $z\sim (4-6)$. Observational data are taken from \citet{Stark2013} (blue circles), \citet{Gonzalez2014} (red triangles), and \citet{Khusanova2020} (green inverted triangles). Our model predictions all assume $\epsilon=0.02$, but different values for $\eta$ and $s$ as indicated in the legend. In particular, the dotted line represents a case with weak outflows and no entrainment, $\eta=2$ and $s=1$, which seems to be a reasonable match to the data. However, models with larger values of $\eta\sim 15$ require entrainment to match the data. Our fiducial model using our mass and redshift dependent values for $s$ and $f_{sa}$ is a very good match to the data, where the data is bracketed by values of $\eta=15$ and $2$ (cyan and magenta lines, respectively).}
    \label{fig:sfr_lowmass}
\end{figure}

In \Cref{fig:sfr_highmass} we focus on halos with $\Mv\sim 10^{12}\msun$ at $z\sim (1.5-4)$, with observations taken from \citet{Pearson2018} and \citet{Leja2022}. In \Cref{fig:sfr_lowmass}, we present results for lower mass halos at higher redshifts, $\Mv\sim 2\times 10^{11}\msun$ at $z\sim (4-6)$, with observations taken from \citet{Stark2013}, \citet{Gonzalez2014}, and \citet{Khusanova2020} for galaxies with corresponding stellar masses given by the stellar-to-halo mass relation from \citet{Behroozi19}. 
For the massive lower-$z$ halos, our model predictions assume a fixed $\eta=2$ as measured in various observations \citep{Newman2012,Hogarth2020,Carniani2023}, although numerical simulations suggest much higher values \citep{Muratov2015,Nelson2019}. The mass loading factor increases with decreasing galaxy stellar mass due to a lower potential well and scales with stellar mass as $\eta\propto M_*^{-1/3}$ for momentum-driven winds \citep{Murray2005, OppenDave08} and $M_*^{-2/3}$ for energy-driven winds \citep{FG2012}, although some numerical simulations show a shallower slope \citep{Oppenheimer2010}. In order to account for this uncertainty, for lower-mass, higher-$z$ halos, we explore the range $\eta=(2-15)$, which roughly brackets the predictions for models with no mass dependence and those with a strong mass dependence. 

We begin by focusing on the massive halos in \Cref{fig:sfr_highmass}. We show results for a fixed entrainment-induced boost factor of $s=3$, $s=2$, and $s=1$. As in \citet{Dekel2014}, we find that the model without entrainment, $s=1$, underpredicts the observed sSFR. The observations seem to favour a boost factor of $s\sim 3$ at $z\sim (1.5-2.5)$, though a lower boost factor at higher redshifts, suggesting that a constant boost factor is not a good fit to the observed data. However, when we consider our full model for a halo mass and redshift-dependent boost factor as described in \Cref{sec:macc_m_z}, we obtain a good fit to the data regardless of whether we assume a self-similar collapse model (cyan line) or an isothermal model (magenta line) for the structure of intergalactic filaments (see \Cref{sec:model_filament}). The main reason for the effective decline in the boost factor at $z\gsim 2.5$ at these halo masses is the predicted drop in the cold gas fraction in cosmic web filaments due to the increase of the filament virial temperature with redshift. 

For the lower mass, higher redshift case presented in \Cref{fig:sfr_lowmass}, we see that a model with no entrainment, $s=1$, provides the lower bound of observations with low feedback with a mass loading factor of $\eta=2$. This is consistent with the results of \citet{Dekel2014} without stellar accretion $f_{sa}\ = 0$, since our model gives a stellar accretion of $0.03$ at these lower masses. 
For this range of mass and redshift, both the CGM and the filaments are partially hot and partially cold, yielding typical values of $s\sim 2$ and $f_{sa}\lsim 0.03$. With the entrainment predicted from our model, the lower and upper bounds of the observations are given by the maximum (cyan) and minimum (magenta) bounds of assumed $\eta$, yielding a good fit to most observational data, though the predicted sSFR declines more steeply towards lower redshift than some of the observational data seem to suggest \citep{Stark2013}.

To summarise, we have shown that a self-consistent model for entrainment of hot CGM gas onto cold streams penetrating massive galaxies, when added to the minimal bathtub toy model presented in \citet{Dekel2014}, can resolve the discrepancy between the observed and predicted values of the sSFR at $z>2$.

\section{Discussion}
\label{sec:disc}


\subsection{Caveats and Additional Physical Effects}
\label{sec:phys}

\smallskip
Our current analysis examines the impact of radiative cooling, self-shielding, and the halo gravitational potential on the evolution of cold streams feeding massive galaxies during cosmic noon and their interaction with the hot CGM. However, to draw firm conclusions regarding the evolution of astrophysical streams, we must consider several additional physical processes that are missing from our current analysis. These include self-gravity, magnetic fields, thermal conduction, turbulence in the CGM and the stream, and the initial penetration of the stream through the accretion shock. In this section, we discuss the potential impact of these processes, which we plan to explore in more detail in future work. 

\smallskip
The impact of \textit{self-gravity} on stream evolution and the stream-CGM interaction has been explored by \citet{A19}, without considering radiative cooling or the halo potential. They found that the self-gravity of the stream can either lead to gravitational fragmentation and collapse if the line-mass is large, or to suppression of KHI due to buoyancy when the line-mass is small. The maximal line-mass for which the stream can maintain hydrostatic equilibrium under its own self-gravity is $\lambda_{\rm max} = a c_s^2/G$, where $a$ is a unitless factor that depends on the adiabatic index of the stream and is $a=2$ for an isothermal stream \citep{Ostriker64}. \citet{M18a} estimated that the cold gas streams feeding galaxies at $z=2$ should have a line-mass on the order of $\lambda\sim \lambda_{\rm max}$. Above this value, the stream will collapse radially and eventually fragment into gravitationally unstable clumps with a separation of the order of a few times the stream diameter \citep{Inutsuka92}, a process which is thought to play an important role in star-formation within filamentary structures in giant molecular clouds (GMCs) in low-$z$ galaxies \citep[e.g.][]{Andre10,Andre14,Arzoumanian11}. Even when $\lambda<\lambda_{\rm max}$, the stream can fragment gravitationally due to long wavelength axisymmetric perturbations \citep{Nagasawa,Hunter97,Hunter98,Heigl16,Heigl18,A19}. Without radiative cooling, the clouds which form as a result of stream fragmentation are sub-Jeans and pressure-confined by the CGM, with their mass approaching the thermal Jeans mass as $\lambda\rightarrow \lambda_{\rm max}$ \citep{A19}. However, when cooling is included, the clouds are expected to become gravitationally unstable and collapse (\citealp{Clarke16,Clarke17}; Aung et al., in preparation). This process can lead to gravitational collapse, star formation in streams within the CGM, and possibly even the formation of globular clusters at high redshift \citep{M18a,Bennett.Sijacki.20}. At very low values of the line-mass, shear between the stream and the CGM prevents fragmentation. At the same time, buoyancy forces stabilise the growth of the shear layer and prevent the stream from fully mixing into the CGM \citep{A19}. If the stream fragments into clumps, this will change the dynamics of the cold gas and the inflow rate because, unlike long cylindrical streams, spherical gas clouds experience ram pressure and additional gas drag, which increase the deceleration and mixing beyond those due only to KHI \citep[e.g.,][]{Forbes2019,Tan23}. However, these effects are likely to be minimal in clouds that form as beads-on-a-string along a fragmented stream, since the interclump medium likely remains co-moving with the clumps. 

\smallskip
\textit{Magnetic fields} have been found to stabilise KHI and suppress the growth of the shear layer in planar, cylindrical, and spherical geometries \citep{Ferrari81,Birkinshaw90,Berlok2019}. This has mostly been studied for non-radiative shear layers. However, relatively few studies to date have explored the evolution of shear layers when considering both radiative cooling and magnetic fields. The evolution is not obvious, since while magnetic fields prevent the mixing of the stream and CGM gas, it is precisely this mixing which leads to entrainment when radiative cooling is considered. Thus, it remains unclear what the net effect will be in terms of the competition between these processes, which will likely depend on the properties of the field. Studies suggest that the result greatly depends on the ratio of thermal to magnetic pressure $\beta\equiv P_{\rm therm}/P_{\rm mag}$ and the direction of the initial magnetic field \citep{Sparre2020,Ledos2023, Bruggen2023,HPineda2023}, and that the answer may be different for spherical clouds compared to planar shear layers \citep{Ji2019,GronkeOh20,Li20}. 
Early results for cylindrical streams suggest that entrainment is suppressed for initial values of the ratio of thermal to magnetic pressure $\beta\lsim 1000$, while for higher values of $\beta$ the evolution is very similar to the hydrodynamic case (Mandelker et al., in preparation). Typical $\beta$ values for cold streams near the virial radii of massive halos at high-$z$ are uncertain, but cosmological simulations suggest that they could be in the range of $\beta\sim (10^2-10^6)$ \citep{Pakmor.etal.2020,Lu2023}. 

\smallskip
The mixing and cooling of gas in the shear layer can be hindered if there is efficient \textit{thermal conduction}. Based on previous work by \citet{Begelman1990_cond} and \citet{Armillotta2016}, \citetalias{M20a} argued that thermal conduction should be important for the evolution of cold streams if the stream radius, $R_{\rm s}$, is smaller than the Field length, given by 
\begin{equation}
\label{eq:field_length_final}
L_{\rm field} \simeq 0.2 \kpc\frac{\delta_{100}^{7/4} T_4^{7/4}}{\Lambda_{-23}^{1/2} n_{\rm s, 0.01}},
\end{equation}
where $T_4 = T_s/10^4\K$, $\delta_{100}=\delta/100$, $n_{\rm s,0.01}=n_{\rm H,0}/0.01\cmc$, and  $\Lambda_{-23}=\Lambda/10^{-23}{\rm erg}\cmc\,{\rm s}^{-1}$ is the cooling rate normalised to the cooling rate at $1.5 T_s$ in the presence of the UV background at $z=2$. \citetalias{M20a} further deduced that $L_{\rm field}$ is comparable to the critical stream radius above which entrainment occurs, $R_{\rm s, crit}$. Thus, whenever the cooling time is shorter than the stream disruption time, thermal conduction is ineffective in smoothing the stream. However, comparing the width of the mixing layer to the Field length yields an intermediate regime where thermal conduction can smooth out the mixing layer but not the stream itself \citep{Bruggen2023}, thus slightly suppressing the mass entrainment rate and Lyman-$\alpha$ luminosity \citep{Ledos2023}. We note that these estimates are based on the properties of streams at the virial radius, while the stream becomes narrower and denser at smaller halocentric radii. As we have seen, the stream radius roughly scales as $R_s \propto n^{-1/2}$ (\equsnp{nhalo}-\equmnp{rstream_rhalo}), while $L_{\rm Field} \propto n^{-1}$ (\equnp{field_length_final}). Therefore, if thermal conduction is ineffective at the virial radius, it should be ineffective in the inner CGM as well, similar to our arguments about the effectiveness of entrainment in the inner versus outer halo. 

\smallskip
Our analysis has assumed that the streams are in thermal pressure equilibrium with the CGM, neglecting non-thermal sources of support such as \textit{turbulence or vorticity}. However, cosmological simulations suggest that cosmic web filaments streams are highly turbulent and strongly supported by rotation and vorticity even prior to penetrating the virial shock around massive halos \citep{Codis12,Codis15,Laigle15,Lu2023}. Furthermore, theoretical studies of filament growth through radial accretion show that the accretion causes turbulence to build up inside the filament and contribute to its support, with typical Mach numbers of order unity \citep{Heitsch13,Clarke16,Clarke17,Heigl18b,M18a}. Non-thermal motions in the form of turbulence and rotation are also important in the CGM itself \citep{Oppenheimer2018, Lochhaas2021, Lochhaas2023}. While turbulence can suppress the growth of cold gas through radiative mixing layers, it can also enhance it under certain circumstances due to the larger surface areas induced by the fractal geometry of the turbulence \citep{Gronke2022}. Such non-thermal effects must be carefully considered in future work in order to describe stream evolution.

As the cold gas stream enters the CGM, it passes through the \textit{virial accretion shock} surrounding the dark matter halo. If the cold stream were to also get shocked at the halo virial radius, the increased pressure can lead to its expansion and eventual disruption \citep{db06,Cornuault2018}. However, cosmological simulations suggest that filaments do not themselves experience a head-on shock at the virial radius \citep{Bennett.Sijacki.20}, with such a shock appearing only at $r\sim 0.3\Rv$ \citep{Zinger2018}. On the other hand, the confining pressure around the cold stream increases by an order of magnitude or more as it penetrates the shock-heated CGM \citep{Lu2023}. This may cause the stream to `shatter' into tiny cloudlets, similar to the effects seen in spherical clouds undergoing a large sudden increase in confining pressure \citep{McCourt.etal.2018,GronkeOh2018,Banda21}. A further complication may arise from the interaction of streams with galactic winds induced by supernova or AGN feedback in the ISM. These effects should be explored in detail in future work, using both idealised simulations of stream-shock interactions, and fully cosmological simulations with enhanced spatial refinement on streams that will allow these processes to be resolved \citep[see, e.g.][]{vdv19,Peeples19,Hummels19,Bennett.Sijacki.20,Mandelker21}. Only then can we draw firm conclusions regarding the evolution of streams in the CGM. 

\subsection{Comparison to Other Models and Simulations}
In this section, we compare two aspects of our results to previous results in the literature. The first relates to the net forces acting on the stream as it flows down the potential well of the dark matter halo, and the second relates to the impact of entrainment on the sSFR of high-$z$ galaxies.

We found that the stream constantly accelerates as it flows from $1.1\Rv$ to $0.1\Rv$ in all cases, but with a net acceleration that is always smaller than free-fall. On the other hand, previous work has shown that cold spherical clouds falling through the hot CGM in an external gravitational potential reach a constant `terminal' velocity, with the inward gravitational acceleration balanced by the deceleration caused by ram pressure and an effective drag force arising from the cold cloud sharing its momentum with the hot gas due to mixing \citep{Tan23}. These effects seem to become dominant only after the first free-fall time, with the clouds close to free-fall at earlier times. In the case of cold streams, we find that the entrainment time scale is much longer than the virial crossing time scale, and the entrained gas is thus not enough to significantly slow the bulk flow of the stream. 
If we were to naively apply the formalism of \citet{Tan23} to the case of cold streams, we would conclude that the streams would reach a terminal velocity of $\sim (5-6)V_{\rm v}$ after $\sim 2$ virial crossing times. This is clearly not self-consistent, as the streams would already have collided with the central galaxy by that point. Thus, cold streams inflowing through an NFW potential are unlikely to ever achieve terminal velocity, despite being somewhat slowed down with respect to the free-fall velocity. 

We found that the observed excess in the sSFR of galaxies at cosmic noon can be explained by the entrainment of hot CGM gas onto the cold stream, which increases the cold gas accretion rate onto the galaxy with respect to the accretion rate onto the dark matter halo. 
An alternate approach was taken by \citet{Mitra2015}. These authors used a similar bathtub model to the one we employ here, but introduced the recycling of gas that was previously ejected from the galaxy due to outflows following some delay time, which was a parameter of their model. Our entrainment scenario also allows the ISM to reaccrete gas which was previously ejected into the CGM, and provides an explicit mechanism for how such recycling may occur. However, we note that the entrained gas is not comprised solely of gas previously ejected from the galaxy, but also of gas accreted onto the halo in the `hot mode', outside of the cold streams. Although our model predicts a final accretion rate similar to the model of \citet{Mitra2015}, a detailed study of the connection between gas recycling and entrainment is left for future work. 

Finally, we note that some cosmological hydrodynamic simulations appear to obtain sSFRs for massive galaxies at $z\sim 2$ that are consistent with observations \citep[e.g.,][]{Nelson2021}. The details of how this is achieved in these simulations, through recycling, entrainment, or some other process, and how these simulations differ from previous studies that underpredicted the sSFR, is not yet clear and should be the focus of future studies.

\section{Summary and Conclusions}
\label{sec:conc}

\smallskip
We study the evolution of cold accretion streams that feed massive star-forming galaxies at $z\sim (2-4)$ from the cosmic web, flowing through their hot CGM towards the central galaxy. These streams are subject to hydrodynamic and thermal instabilities as a result of their interaction with the ambient hot CGM gas. Previous works have shown that this interaction can lead to the entrainment of hot CGM gas onto the cold streams through a radiative turbulent mixing layer \citepalias{M20a}, thus increasing the cold gas accretion rate towards the galaxy \citepalias{M20b}. However, these previous works used simulations that did not include the gravitational potential of the host halo \citepalias{M20a} or analytic arguments which accounted for the halo potential but in an overly simplistic manner and without the corresponding numerical simulations \citepalias{M20b}. Here, we used numerical simulations that include the halo potential along with an improved analytical treatment to study the evolution of the stream. We then incorporate our results into a `bathtub' model for galaxy evolution, based on the minimal bathtub toy model introduced in \citet{Dekel2014}, to explore the effects of CGM entrainment onto cold streams on the SFR histories of galaxies. Our main results can be summarised as follows:

\begin{enumerate}

\smallskip
\item We derive equilibrium configurations for cold streams embedded in a hot CGM in hydrostatic equilibrium within the potential well of an NFW dark matter halo. The streams become denser, narrower, and colder as they accelerate towards the halo centre.

\smallskip
\item When perturbations are introduced at the base of the stream near the virial radius, the streams nonetheless survive to the halo center across all parameters studied, even those that would suggest that the streams should disrupt based on their properties at $\Rv$. Furthermore, the radial mass inflow rate of the cold gas increases by a factor of $2-3.5$ as the stream approaches the centre. This is mainly due to the entrainment of hot CGM gas, which mixes with cold stream gas and cools. The entrainment is stronger for denser streams due to the faster cooling rate, and for initially slower streams due to the longer time the stream spends in the CGM. 

\smallskip
\item The dissipation rate of mechanical energy and thermal enthalpy induced by the stream-CGM interaction as the stream flows down the potential well of the host halo is sufficient to power observed Ly$\alpha$ blobs with $L\gsim 10^{42}{\rm erg\,s}^{-1}$ for certain stream parameters. This is also in agreement with the previous model of \citetalias{M20b}. The energy sources are the infall into the halo gravitational potential well and the radiative cooling in the stream-CGM mixing layer, combined with UV fluorescence. 

\smallskip
\item The increase in the radial mass inflow rate of the cold gas and the dissipation rate stays the same whether or not self-shielding of dense gas from the UV background is considered, with the only difference being the fraction of the emission that is sourced from the stream-CGM interaction versus the UV background.

\smallskip
\item The simulation results are well described by an analytic model that improves upon the model proposed in \citetalias{M20b} by accounting for realistic density and temperature profiles in the hot CGM and cold streams. Based on the model and our assumed fiducial stream properties, we predict that the cold gas accretion rate onto galaxies that live in the centres of dark matter halos with $\Mv\gtrsim 10^{12}\msun$ at $z\sim (2-3)$ is $2-3.5$ times higher than the cosmological accretion rate at the virial radius of their dark matter halos. The enhancement in the inflow rate decreases towards lower redshifts due to the lower densities and slower cooling at later times.

\smallskip
\item  Two necessary conditions for the increase in the cold gas accretion rate are (1) the existence of a hot CGM in the host dark mater halo, and (2) that the intergalactic cosmic web filaments contain a substantial amount of cold gas in their cores before they enter the halo. Using these conditions, we extend our model to predict the cold gas accretion rate onto galaxies as a function of halo mass and redshift and identify the regimes where such conditions are satisfied, roughly $\Mv\sim (10^{11.5}-10^{12.5})\Msun$ at $z=(1-5)$. 

\smallskip
\item Using an analytic bathtub toy model, we compute the star formation rates of galaxies as a function of halo mass and redshift, accounting for the boost in the cold gas accretion rat. Our model predictions agree with the observed star formation rate of massive star-forming galaxies at $z=2-5$.

\end{enumerate}

\section*{Acknowledgments} 
The simulations were carried out on the Stampede2 cluster through the ACCESS grant PHY210069 and the High Performance Computing clusters at the Hebrew University Research Computing Services and Yale Center for Research Computing. 
HA acknowledges support from the Zuckerman Postdoctoral Scholar Program. NM is supported by the Israel Science Foundation (ISF) grant 3061/21 and U.S-Israel Binational Science Foundation (BSF) grant 2020302. AD is supported by ISF grant 861/20. DN and FvdB are supported by the National Science Foundation (NSF) through grants AST-2307280. This research was supported in part by grant NSF PHY-1748958 to the Kavli Institute for Theoretical Physics (KITP). We acknowledge Keshav Raghavan for his contributions on the self-similar profile of filaments. We thank the anonymous referee for their helpful suggestions and comments on
the manuscript, and Andrea Ferrara and Corentin Cadiou for the discussions on earlier version of the research.  We also acknowledge the use of the following python packages numpy \citep{numpy}, astropy \citep{astropy}, COLOSSUS \citep{colossus} and matplotlib \citep{matplotlib} for plotting and data analysis.

\section*{Data Availability} 
The data from idealised simulations will be shared on request. The analytic model used in the study can be obtained at \href{https://github.com/h-aung/filament_sfr_model}{https://github.com/h-aung/filament\_sfr\_model}.


\bibliography{biblio}
\bibliographystyle{mnras}

\appendix

\section{Hydrostatic Profile for the Hot CGM}
\label{app:Hyd_CGM}

In this section, we provide the solution to \equ{Hydrostatic} for the density profile of the hot CGM in hydrostatic equilibrium within the gravitational potential of an NFW halo, as described in \Cref{sec:hydrostatic}. The enclosed mass profile is 
\be 
\label{eq:NFW_mass}
M(r)=\Mv\frac{f(cx)}{f(c)},
\ee
where $\Mv$ is the halo virial mass, namely the total mass enclosed within the virial radius, $\Rv$, $x\equiv r/\Rv$, and $c$ is the halo concentration. The function $f(x)$ is given by 
\be 
\label{eq:fc}
f(x)={\rm ln}\left(1+x\right)-\frac{x}{1+x}.
\ee

Inserting this into \equ{Hydrostatic} along with the assumption of a polytropic gas profile, namely 
\be 
\label{eq: polytrope}
P_{\rm h}(r)=P_{\rm h,0}\left(\frac{\rho_{\rm h}(r)}{\rho_{\rm h,0}}\right)^{\gamma'}, 
\ee 
with $P_{\rm h,0}$ and $\rho_{\rm h,0}$ the CGM pressure and density at some radius $r_0$, we obtain the following solution for the density, 
\be 
\label{eq:CGM_density}
\frac{\rho_{\rm h}(r)}{\rho_{\rm h,0}} = \left[1+\eta\left(\frac{{\rm ln}(1+cx)}{x} - \frac{{\rm ln}(1+cx_0)}{x_0}\right)\right]^{\frac{1}{\gamma'-1}},
\ee 
where $x_0=r_0/\Rv$ and 
\be 
\label{eq:prof_fac}
\eta \equiv \frac{\gamma'-1}{\gamma'}\,\frac{G\Mv/\Rv}{P_{\rm h,0}/\rho_{\rm h,0}}\,\frac{1}{f(c)}.
\ee 

We see that the density profile is cored, approaching a constant at small radii. Following \citet{KS01}, we demand that the gas density profile follow the NFW profile outside of the core, over a wide range of radii. This condition sets the parameters $\gamma'$ and $(G\Mv/\Rv)/(P_{\rm h,0}/\rho_{\rm h,0})$ as a function of $c$. For our fiducial value of $c=10$, \citet{KS01} obtain $\gamma'\sim 1.185$ and $(G\Mv/\Rv)/(P_{\rm h,0}/\rho_{\rm h,0})\sim 3.536$. The latter corresponds to a ratio of virial velocity to adiabatic sound speed at $r_0$ of $\Vv/c_{\rm b}=[(G\Mv/\Rv)/(\gamma P_{\rm h,0}/\rho_{\rm h,0})]^{1/2}\sim 1.45$. 
In our simulations, we set $r_0=1.1\Rv$ and vary $\rho_{\rm h,0}=\rho_{\rm s,0}/\delta_0$ (see \Cref{tab:sim}). 

\section{Self-similar Density Profiles of Filaments}
\label{app:self-similar}

The self-similar filament profiles are derived by combining elements of the collisionless cylindrical collapse model of \citet{FG84} and the collisional spherical collapse model of \citet{bertschinger1985}. Following \citet{FG84}, the matter will break away from the expanding background and collapse at a turnaround radius if the gravitational pull due to the enclosed density is large enough to overcome the Hubble flow. Following \citet{bertschinger1985}, we do not assume virialization after the infalling mass shell reaches a certain radius, but rather consider shell crossing for dark matter and the formation of an accretion shock for gas. We assume an Einstein-de Sitter universe with $\Omega_m=1$, $a\propto t^{2/3}$ and a background matter density $\rho_b = 1/6\pi G t^2$, with $t$ the cosmic time. The model also assumes that the filament is infinite along its axis. Each mass shell around the overdense cylindrical region reaches its turnaround radius at some time $t_{\rm ita}$, the time of initial turnaround, and then proceeds to fall towards the filament axis. This sets the initial condition for the mass shell, where the turnaround radius is $r_{\rm ta}\equiv r(t_{\rm ita})$, the enclosed line-mass at turnaround is $\Lambda_{\rm ta}\equiv \Lambda(t_{\rm ita})$, and the initial velocity is $v_{\rm ta}\equiv v(t_{\rm ita})=0$. The filament line-mass enclosed within the turnaround radius at time $t$ increases as a function of time as\footnote{Self-similar models for halo mass growth assume $m(r_{\rm ta}) \propto a^{s}$ \citep{FG84,shi2016b}. For filaments, the filament axis is expanding due to the expansion of the universe, which decreases the line-mass and leads to the $-1$ in the power law exponent.} $\Lambda(r_{\rm ta}) \propto a^{s-1}\propto t^{2(s-1)/3}$. Accordingly, the turnaround radius grows as $r_{\rm ta}(t) \propto \sqrt{\Lambda/\rho_b} = a^{3\delta/2}$, where $\delta=2(1+s/2)/3$. The dark matter mass shell then follows the equation of motion \citep{FG84} 
\be
\label{eq:app:dm_motion}
\frac{\rmd^2 r}{\rmd t^2} = \frac{d(Hr)}{dt}-\frac{2G(\Lambda-\Lambda_b)}{r} = \frac{2 \pi G\rho_b r }{3} - \frac{2G\Lambda}{r}, 
\ee
{\no}with $H=a^{-1}da/dt$ the Hubble constant at time $t$, and $\Lambda_b=\pi r^2 \rho_b$.

\begin{table}
\centering
 \begin{tabular}{| c | c | c | c | c | c |} 
 \hline
 $t$ & $r$ & $\Lambda$ & $p$& $v$& $\rho$\\ 
 \hline
 $\xi$ & $\lambda$ & $M$ & $P$& $V$& $D$\\ 
 \hline
 $\ln{\frac{t}{t_{\rm ita}}}$ & $\frac{r}{r_{\rm ta}}$ & $\frac{\Lambda}{\rho_b \pi r_{\rm ta}^2}$ & $\frac{p}{\rho_b (r_{\rm ta}/t)^2}$ & $\frac{v}{r_{\rm ta}/t}$ & $\frac{\rho}{\rho_b}$ \\
 \hline
\end{tabular}
\caption{The normalization assumed under the self-similar model. The top row indicates physical quantity. The middle indicates the corresponding dimensionless quantity. The bottom row indicates the relation between the dimensionless quantity and the physical quantities. }
\label{table:scaling}
\end{table} 

The self-similar model assumes that the filament profile is universal for a given mass accretion rate, and we can, therefore, remove the time dependence when all parameters are normalised by appropriate quantities (see \Cref{table:scaling}). The equation of motion can then be expressed in dimensionless form as
\begin{equation}
\frac{\rmd^2\lambda}{\rmd \xi^2} + (2\delta-1)\frac{\rmd\lambda}{\rmd \xi} + \delta(\delta-1)\lambda = \frac{\lambda}{9}-\frac{M}{3 \lambda}. 
\label{eq:app:dm_norm_motion}
\end{equation}

We solve \equ{app:dm_norm_motion} iteratively as follows. We begin by assuming a power-law mass profile, and then numerically integrate the equation with the outer boundary condition $M(\lambda=1) = M_{\textrm{ta}}$ and $\mathrm{d}\lambda / \mathrm{d}\xi (\lambda=1) = -\delta$. Once we obtain the solution for the trajectory of the mass shell, we update the total enclosed line-mass at radius $\lambda$ according to \citet{Bertschinger1985_void, bertschinger1985}
\begin{equation}
    M(\lambda) = M_{\textrm{ta}} \sum_{i=1}^{N(\lambda)} (-1)^{i-1} \exp(-2 (s-1) \xi_i/3),
\label{eq:app:dm_mass}
\end{equation}
where the index $i$ runs over all $N(\lambda)$ mass shells that are currently at radius $\lambda$, and $\xi_i$ are the times with respect to each mass shell's turnaround time. $M_{\rm ta}$ is the normalised line mass inside the current turnaround radius. Thus, $M_{\rm ta}{\rm exp}[-2 (s-1) \xi_i/3]=M_{\rm ta}(a_{\rm ita}/a_i)^{s-1}=\Lambda_{\rm ita}/(\pi r_{\rm ta}^2 \rho_{\rm b})$ is the enclosed line-mass when the shell was at the turnaround normalised by the current density and the turnaround radius, so that the overall normalisation of the mass profile is uniform among all shells. 
The alternating signs $(-1)^{i-1}$ account for the fact that the shells at the radius $\lambda$ are alternating whether they are flowing in or out. The first shell is on its way in along the first infall, the second has fallen in and is on its way back out towards the first splashback, the third is on its way in along the second infall, etc. 
We then insert the updated mass profile obtained with \equ{app:dm_mass} back into \equ{app:dm_norm_motion} and solve it again to obtain a new mass profile. We repeat this process until the mass profiles have converged to within $<3\%$.

The collisional gas, on the other hand, follows the continuity equations expressed as
\begin{align}   
\frac{\rmd\rho}{\rmd t} &= -\frac{\rho}{r} \frac{\partial}{\partial r}(rv), \nonumber\\ 
\frac{\rmd v}{\rmd t} &= \frac{2 \pi G}{3} \rho_b r  - \frac{2 G\Lambda}{r} + \frac{1}{\rho} \frac{\partial p}{\partial r}, \nonumber\\
\frac{\rmd k}{\rmd t} &= 0, \nonumber\\
\frac{\partial \Lambda}{\partial r} &= 2\pi r\rho, \label{eq:app:gas} 
\end{align}
{\no}with the entropy $k\equiv p \rho^{-\gamma}$, and the Lagrangian derivative ${\rm d}f/{\rm d}t\equiv \partial f/\partial t+{\bf{v}}\cdot {\bf{\nabla}}f=\partial f/\partial t+v\,\partial f/\partial r$.
The gas is assumed to be pressureless outside the shock and infalls similarly to dark matter until it is shock-heated at a radius $r_{\rm sh}$. The post-shock properties are given by 
\begin{align}  
v_2 &= \frac{\gamma-1}{\gamma+1} \left[v_1 - v_{\rm sh}\right] + v_{\rm sh}, \nonumber\\ \rho_2 &= \frac{\gamma+1}{\gamma-1} \rho_1, \nonumber\\ p_2 &= \frac{2}{\gamma+1}\rho_1 \left[v_1 - v_{\rm sh} \right]^2, \nonumber\\ \Lambda_2 &= \Lambda_1, \label{eq:app:shock}
\end{align}
assuming infinite Mach number due to pressureless pre-shock conditions. $v_{\rm sh}$ is the speed at which the accretion shock propagates, given by differentiating $r_{\rm sh}(t)= \lambda_{\rm sh} r_{\rm ta}(t)$, where $\lambda_{\rm sh}$ is constant with time due to the assumption of self-similarity. The continuity equations and the shock jump conditions can be rewritten as follows. 
\begin{align}
-2D + \left(V - \delta \lambda\right)D'  &=-\frac{D}{\lambda}(\lambda V)', \nonumber\\ 
V(\delta-1) + (V-\lambda\delta)V' &= \left[\frac{\lambda}{9} - \frac{M}{3\lambda}\right] - \frac{P'}{D}, \nonumber \\
(V - \lambda \delta)\frac{(PD^{-\gamma})'}{PD^{-\gamma}} &= 2(1-\gamma) +2(1-\delta),\nonumber \\ 
M' &= 2 \lambda D, \label{eq:app:norm_gas}
\end{align}
and
\begin{align}  
V_2 &= \frac{\gamma-1}{\gamma+1} \left[V_1 - \lambda_{\text{sh}} \delta\right] + \lambda_{\text{sh}} \delta, \nonumber\\ 
D_2 &= \frac{\gamma+1}{\gamma-1} D_1, \nonumber\\ 
P_2 &= \frac{2}{\gamma+1}D_1 \left[V_1 - \lambda_{\text{sh}} \delta \right]^2, \nonumber\\ 
M_2 &= M_1, 
\label{eq:app:norm_shock}
\end{align}
where $\lambda_{\rm sh}$ is the normalized shock radius. These equations are solved using the same initial conditions at the turnaround radius as for dark matter, and the shock radius is set such that the solution ensures the inner boundary condition $V=0$.

Note that the model assumes a matter-dominated EdS universe, which is valid for $z\gtrsim 2$. The accretion rate $s$ for the filament can be estimated by differentiating \cref{eq:Lambda_general} with time and calculating $\rmd \log M/\rmd \log a$. This gives $s\approx 1.2$, assumed throughout our model. Given the redshift of the halo and the line-mass of the filament from \cref{eq:Lambda_general} and the specific $s$ of the filament, we can calculate $r_{\rm ta}$ to provide the normalisation of the filament profile.

\section{Resolution Convergence}
\label{app:resolution}
\begin{figure}
    \centering
    \includegraphics[width=0.49\textwidth]{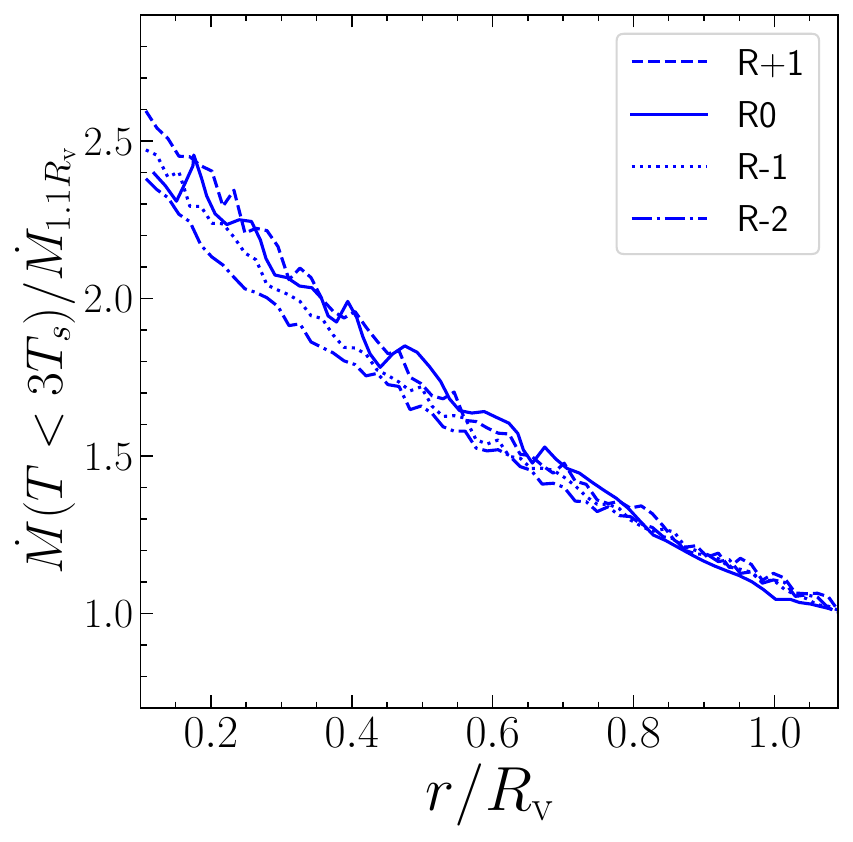}\\
    \includegraphics[width=0.49\textwidth]{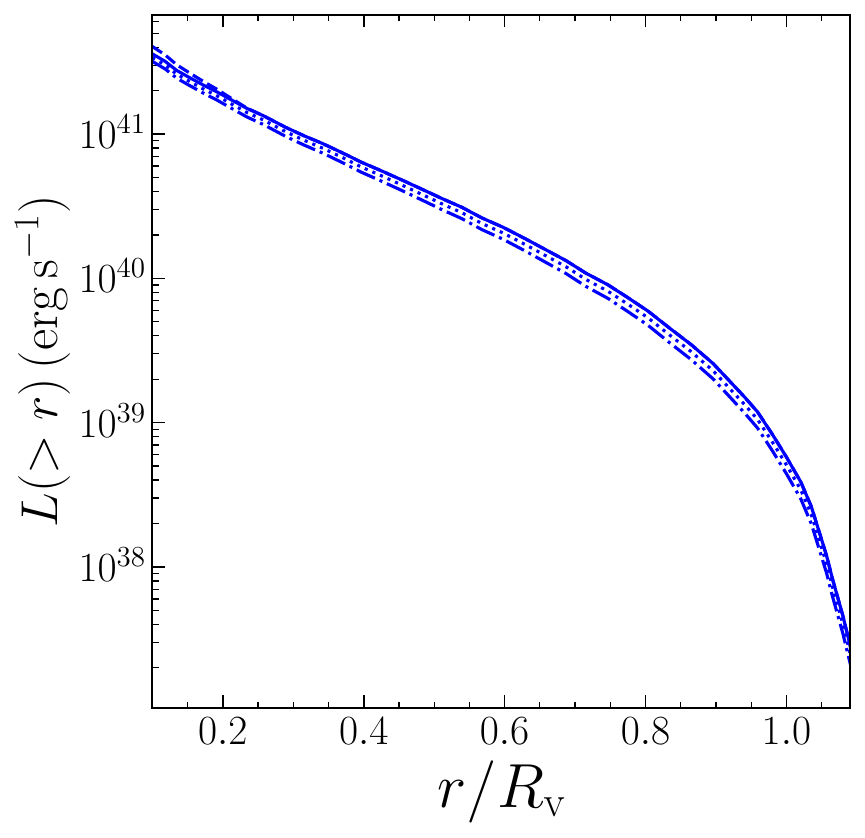}
    \caption{
    The cold gas mass inflow rate (top) and the integrated Lyman-$\alpha$ luminosity profile (bottom) as a function of halocentric radius, $\rhalo$, in simulations with different resolutions for our fiducial parameters (first row in \Cref{tab:sim}). We compare the result of $R0$, the fiducial resolution, to two lower resolution simulations ($R-1$ and $R-2$) and one higher resolution simulation ($R+1$). There is a mild trend of increased inflow rate and luminosity with higher resolution, but the differences are extremely small, and all simulations converge to within $<5\%$.
    }
    \label{fig:reso_profiles}
\end{figure}

We repeat the fiducial simulation of $V_{s,0}=V_v, \delta_0 = 100, n_{H,0}=0.01\cmc$ (first row in \Cref{tab:sim}), with 2 lower resolution simulations (denoted $R-1$ and $R-2$) and 1 higher resolution simulation ($R+1$). In the lower-resolution simulations, we increase cell sizes in all regions by a factor of 2 and 4 for $R-1$ and $R-2$, respectively. For $R+1$ simulation, we add an additional refinement region, resolving the region of $z<0.225\Rv$ and ${\rm max}(|x|,|y|)<0.25\Rso$ by another factor of 2. In \Cref{fig:reso_profiles}, we show that the profiles of mass inflow rate and Lyman-$\alpha$ luminosity generally converge, with a very slight decrease in entrainment rate as the resolution gets lower. This is in agreement with previous studies of cold spherical clouds in hot wind tunnels, where the cold gas entrainment rate is converged for resolution as low as 8 cells per cloud radius, which would be equivalent to $R-3$ 
\citep{GronkeOh20}.

\end{document}